\documentclass[12pt]{article}

\begin{document}

\catcode`\@=11
\def\lesssim{\mathrel{\mathpalette\vereq<}}
\def\gtrsim{\mathrel{\mathpalette\vereq>}}
\def\vereq#1#2{\lower3pt\vbox{\baselineskip0pt \lineskip0pt
\ialign{$\m@th#1\hfill##\hfill$\crcr#2\crcr\sim\crcr}}}
\catcode`\@=12

\def\alt{\lesssim}
\def\agt{\gtrsim}

\newcommand{\noe}{\hbox{E\kern-.6em\hbox{/}\kern.3em}}

\begin{center}
{\Large TASI lectures: weak scale supersymmetry\\ 
--- a top-motivated-bottom-up approach}\\
\vspace{.2in}
%\author{Gordon L. Kane}
Gordon L. Kane\\
Michigan Center for Theoretical Physics, Randall Physics
Lab,\\
University of Michigan\\
E-mail: gkane@umich.edu
\end{center}
%\maketitle
\begin{center}
CONTENTS
\end{center}

$\circ$ Introduction, Perspective -- since particle physics beyond the SM is
presently in an incoherent state, with lots of static, a long introduction is
needed, including some history of the supersymmetry revolutions, physics not
described by the SM, indirect evidence for low energy supersymmetry
and how flavor physics should be approached.

$\circ$ Derive the supersymmetric Lagrangian -- the superpotential W

$\circ$ Soft supersymmetry breaking -- underlying physics -- $L_{soft}$
-- the MSSM

$\circ$ The $\mu$ opportunity -- R-parity conservation

$\circ$ Count of parameters -- constraints -- measuring the parameters

$\circ$ Connecting the weak and unification scale

$\circ$ Derivation of the Higgs mechanism -- in what sense does supersymmetry
\textit{explain} the Higgs physics

$\circ$ The Higgs spectrum -- $\tan\beta,$ Yukawa couplings, constraints

$\circ$ LEP Higgs physics -- Tevatron Higgs physics can confirm the
Higgs mechanism and coupling proportional to mass -- Higgs sector measurements

$\circ$ \~{g}, \~{N}, \~{C} -- cannot in general measure $\tan\beta$
at hadron colliders

$\circ$ Effects of soft phases -- all observables, not only CPV ones,
$g_{\mu }-2$, EDMs, $\tilde{g}$ phase, LSP CDM, possible connections
to stringy physics

$\circ$ Phase structure of simple D-brane models

$\circ$ Tevatron superpartner searches, signatures

$\circ$ Extensions of the MSSM

$\circ$ The importance of low scale supersymmetry is not only that we
learn of another profound aspect of our world, but also to provide a
window to Planck scale physics, in order to connect string theory and our world

\newpage

\section{Introduction}

For about 400 years we have improved our understanding of the physical world
until we discovered and tested the Standard Model (SM) of particle physics,
which provides a complete description of our world, of all that we see.  We
know that the basic constituents are quarks and leptons, and we have a complete
theory of the strong, weak, and electromagnetic forces.  

We also know that much is unexplained, such as what is the cold dark matter
(CDM) of the universe, and why is the universe made of matter rather than
being an equal mixture of matter and antimatter, neither of which can be
explained by the SM.  Below I will make a longer list of questions the SM
cannot answer.  And there are conceptual reasons also why we expect to find
new physics beyond the SM.  

While a number of approaches to physics beyond the SM have been worth
considering, only one so far has actually explained and predicted phenomena
beyond the SM, the supersymmetric extension of the SM, which will be the focus
of these lectures.  As we will see, the supersymmetric SM has a number of
successes, and as yet no failures.  It is not yet a complete theory in the
sense that we do not yet understand fully the physics of all of its
parameters, but it is a complete effective theory because we can write the
full effective Lagrangian of the theory.  One of its important successes is
that it can be a valid theory to very high energy scales or very short
distances, near to the Planck scale.  Another is that it is not
sensitive to new physics at some high scale. 

Superstring theories have become very attractive in recent years as
well.  They are formulated near the Planck scale with ten dimensions and
presumably unbroken supersymmetry.  String theories only predict or
explain that there is a gravitational force, and that we live in at most
10 dimensions.  What is exciting about them from our point of view is
that they seem to be able to accommodate the SM forces and quarks and
leptons, and possibly explain how these forces and particles originate.
So in these lectures we will assume that the basic theory is a string
theory at the Planck scale (loosely speaking).  We do not distinguish
string theory from M-theory for our top-motivated purposes, since the
effective low scale theory from both will be parameterized by the same
Lagrangian.

Historically, physics has progressed by one basic method, with
experiment and theory intermingling as each level of the world was
understood.  That method will continue to work as supersymmetry is
established experimentally, as its parameters (masses, flavor rotation
angles, phases, vacuum expectation values) are measured, and as
supersymmetry breaking is studied.  But the historical approach can only
take us to a broken supersymmetric theory near the unification and
string scales.  It cannot be used to learn the form of the 10D
supersymmetric string theory.  There is a barrier that can only be
crossed by human imagination.  To cross it we must know the Lagrangian
of the broken supersymmetric theory near the unification and string
scales, and we must understand the 10 D supersymmetric string theory
very well.  Then it will be possible to guess how to jump the barrier.
In my view it will not be possible to do that until the broken
supersymmetric theory near the Planck scale is known.  No amount of
thinking will tell us how to compactify the string theory, or to break
supersymmetry, or to recognize the vacuum of the theory, because there
is no practical way to recognize if one has it right.

Let's pursue this in a little more detail.  Sometimes people argue that
calculating fermion masses will be a convincing way to learn when a
compactification is correct.  But the hierarchy of fermion masses
implies that will be very hard to do.  The small masses are unlikely to
arise at the tree level, but rather depend on non-renormalizable
operators and possibly on supersymmetry breaking effects.  So perhaps
the large masses can be calculated, but not the smaller ones, and if the
large ones have Yukawa couplings of order unity that will be common to
many theories.  It is of course known that huge numbers of manifolds
give three families of chiral fermions.  A little thought suggests very
strongly that most of the usual ``string phenomenology'' is of a similar
nature, and is very unlikely to point toward the correct vacuum.
Indeed, suppose some string theorist already knew how to compactify and
to break supersymmetry and to find the correct vacuum.  How would they
convince themselves, or anyone else?

However, the supersymmetry soft-breaking Lagrangian, $L_{soft}$ may
offer more hope for testing theory. The parameters of $L_{soft}$ are
measurable, though little has been known until recently about how to
measure most of them, and much of these lectures will be about how to
measure them.  If a theorist has an approach to compactification and to
breaking supersymmetry, then $L_{soft}$ is likely to be calculable more
easily than the full Yukawa matrix in that approach, and thus knowledge
of $L_{soft}$ may test ideas better than knowledge of the fermion
masses.  The parameters of $L_{soft}$ may be less sensitive to uncertain
higher order corrections (unless the leading term vanishes in which case
the one-loop radiative correction is usually not hard to work out).  The
flavor structure of $L_{soft}$ depends on the flavor structure of the
Yukawas and may help untangle that.  Progress will come from measuring
$L_{soft}$ at the weak scale, and extrapolating it to the unification
scale.  The patterns of the soft parameters may be typical of one
approach or another to compactification or supersymmetry breaking or
the vacuum structure, so the measured $L_{soft}$ may focus attention
toward particular solutions to these problems.  Superpartners should be
directly detected in the next few years, and once the initial excitement
is past we will turn to the challenging and delightful opportunity to
untangle the data and measure the Lagrangian.

Supersymmetry is an idea as old as the SM, and it has not been the
most fashionable way to describe the real world in recent years.
Consequently many students have not become familiar with supersymmetry
as a practical theory, nor have they seen the arguments for its
validity.  Once superpartners are directly observed it will not be
necessary to include these arguments, but at the present time there is
some static in the messages most students get, so it is worthwhile to
include some tables summarizing why classic supersymmetry is the best
approach.  In these largely pedagogical lectures I will also not focus
on extensive referencing, with apologies to many authors.  Some
references are given to help the reader find the relevant additional
papers.  Many topics are integrated into these lectures, and most have
been worked on by many authors, so I either have to provide extensive
referencing or little referencing, and the latter seems reasonable
here in a pedagogical treatment.  For thorough referencing to the
literature before the past three years the chapters in ref.\cite{1} are
useful.  I will in places follow the approach of Martin in ref. \cite{1},
and he has very good referencing; the larger version of his chapter on
the web is more valuable than the printed chapter \cite{2}.

It is good to recall some of the history of supersymmetry.  We can
basically split it into five ``revolutions'':

\vspace{.2in}

\noindent HISTORY OF SUPERSYMMETRY REVOLUTIONS

\vspace{.1in}

\begin{tabular}
[c]{lll}%
1$^{st}$ & 1970-72 & \ \ \ \ \ \ \ \ \ \ \ \ \ \ \ \ \ \ \ \ The idea\\
2$^{nd}$ & \ \ 1974 & Supersymmetric relativistic quantum theory\\
3$^{rd}$ & \ \ 1975 & \ \ \ Local supersymmetry, supergravity\\
4$^{th}$ & 1979-83 & Supersymmetry solves many problems\\
5$^{th}$ & 2000-03 & Higgs boson and superpartners observed
\end{tabular}

\vspace{.2in}

Next let us consider a list of important questions that the SM does not deal
with.  Consequently, these can point the way beyond the SM.

\section{Physics not described by the Standard Model}
$\bullet$ Gravity

\noindent $\bullet$ Cosmological Constant

\noindent $\bullet$ Dark Energy

\noindent $\bullet$ What is (are) the inflation(s)?

\noindent $\bullet$ Strong CP problem

\noindent $\bullet$ Hierarchy problem

\noindent $\bullet$ How is the electroweak symmetry broken (EWSB)?

\noindent $\bullet$ Gauge coupling unification

\noindent $\bullet$ Matter asymmetry of the universe

\noindent $\bullet$ Cold dark matter

\noindent $\bullet$ 3 families

\noindent $\bullet$ Neutrino masses

\noindent $\bullet$ Values of quark and charged lepton masses

\noindent $\bullet$ Approximate Yukawa unification of bottom, tau, and
perhaps top

\noindent $\bullet$ The value of the Higgs boson mass 115 GeV if the
LEP signal is confirmed

\vspace{.2in}

The SM \textit{cannot} account for or explain any of these.  It can
accommodate some of them.  Any approach that claims to be making any
progress (such as large extra dimension ideas) should be able to deal
with some or most of these simultaneously.  Where do they lead us?
Supersymmetry of the form we are focusing on in these lectures is
relevant to most or all of these.  (There are additional reported
deviations from the SM that could be relevant and arise from
superpartner loops (a) the condition for charged current universality,
or unitarity of the CKM matrix \cite{3}, and (b) the number of neutrinos
is slightly less than 3 \cite{4}.)

\section{The Hierarchy Problem}

The hierarchy problem is the SM problem that quantum corrections raise the
Higgs boson mass up to the highest mass scale there is.  It is a serious
problem --- as someone said, the quantum corrections are not only infinite,
they are large.  The high mass scales do not have to couple directly to the
Higgs boson; the coupling can be through several loops, as Martin explains in
some detail.  All SM masses (W and Z and quarks and charged leptons) are
proportional to m$_{h}$ so if m$_{h}$ is large they are too.  

Supersymmetry was not invented or designed to solve this problem
(contrary to what is often stated), but it did.  If supersymmetry is
unbroken then loops with particles cancel loops with their superpartners
in general.  For broken supersymmetry the effect is proportional to a
power of some couplings times the square of the difference of the masses
of superpartner pairs, and a log of mass ratios.  Any solution of the
hierarchy problem must be insensitive to high scales, and to higher
order corrections.  If an approach is claimed to deal with the hierarchy
problem it must explain how the weak and gravitational scales are
determined.  Later when we discuss EWSB we examine in what sense
supersymmetry provides this explanation.  Sometimes people make
connections between the cosmological constant problem and the Higgs
hierarchy problem, but they are not the same because the calculation
of the cosmological constant sums over all states, while the calculation
of the Higgs mass only sums over states with SM gauge quantum numbers.
Another way to think of the supersymmetry solution is that the Higgs
doublet becomes a chiral supermultiplet so h and its superpartner have
the same mass, and the fermion masses are not quadratically divergent so
its superpartner mass is not quadratically divergent.

\section{Gauge Coupling Unification}

One of the most important things we have learned from LEP is that the
gauge couplings unify at an energy above about $10^{16}$ GeV in a
world described by a supersymmetric theory, though not in the SM. 
Further, where they meet points toward a unification with gravity near
the Planck scale.  Together these make one of the strongest
indications of the validity of the view of physics at the foundation
of these lectures.  Any other view has to claim this unification is a
coincidence!  The gauge coupling unification implies two important
results:

(1) The underlying theory is perturbative up to the unification scale.
 Sometimes it is said there should be a desert (apart from the
superpartners) but that is not so --- only that whatever is in that
range (such as right handed neutrinos) does not destroy the
perturbativity of the theory.

(2) Physics is simpler at or near the unification scale.  That need
not have happened --- nothing in the SM implies such a result. 

There is another important clue.  The supersymmetric gauge coupling
unification misses by about 10\%.  More precisely, the experimental
value of the strong coupling $\alpha_{3}$ is about 10-15\% lower than
the value computed by running down theoretically from the point were
the SU(2) and U(1) couplings meet.  The details are interesting here
--- the one-loop result is somewhat small because of a cancellation,
and the two-loop contribution therefore not negligible.  If one only
took into account the one-loop effect the theoretical value would be
close to the experimental one but the two-loop effect increases the
separation.  Nature is kind here, on the one hand giving us
information about the need for supersymmetric unification, and on the
other giving us a further clue about the physics near the unification
scale, or about particles that occur in the ``desert'' and change the
running somewhat.

\section{(Indirect) Evidence for weak scale supersymmetry}

We have described two of the strongest pieces of evidence for weak
scale supersymmetry.  The third and to some the strongest is that
this approach can explain the central problem of how the electroweak
symmetry is broken --- we will consider that in great detail after we
derive the supersymmetric Lagrangian.  First we list here additional
evidence for weak scale supersymmetry.  Sometimes people wrongly
imagine that supersymmetry was invented to explain some of what it
explains so the approximate date when it was realized that each of
these pieces of evidence existed is listed.  Of course the theory
existed even before it was realized that it solved these problems ---
it was not invented for any of them.  For completeness we include the
evidence we have already examined.

\vspace{.1in}

1980 --- Can stabilize hierarchy of mass scales.

1982 --- Provides an explanation for the Higgs mechanism.

1982 --- Gauge coupling unification.

1982 --- Provides cold dark matter candidate.

1982 --- Heavy top quark predicted.

1992 --- Can explain the baryon asymmetry of the universe.

1993 --- Higgs boson must be light in general supersymmetric theory.

1990 --- Realization that either superpartners are light enough to
find at LEP, or their effects  on precision data must be very small
and unlikely to be observed at LEP/SLC.  Supersymmetry effects at
lower energies arise only from loops, which explains why the SM works
so well even though it is incomplete.

1982/1995 --- Starting from a high scale with a value for
$\sin^{2}\theta_{W}$ of 3/8, which arises in any theory with a unified
gauge group that contains SU(5), and also in a variety of string-based
theories, the value for $\sin ^{2}\theta_{W}$ at the weak scale is
$0.2315$ and agrees very accurately with the measured value.

\vspace{.1in}

Some of these successes are explanations, and some are  correct
predictions.   It is also very important that all of them are
simultaneously achieved --- often efforts to deal with the real world
can apparently work for one effect, but cannot describe the range of
phenomena we know.

There are theoretical motivations for low energy supersymmetry too.  It
is the last four-dimensional space-time symmetry not yet known to be
realized in some way in nature, it adds a fermionic or quantum structure
to space-time, it allows theories to be extrapolated to near the Planck
scale where they can be related to gravity, local supersymmetry is
supergravity which suggests a connection of the supersymmetric SM to
gravity, it allows many problems in string theory and string field
theory to be solved, including stabilizing the string vacuum.  It is
expected, though not yet demonstrated, that low energy supersymmetry is
implied by string theory.  Not all of these necessarily require low
energy supersymmetry.  In any case, improving the theory is nice but is
not strong motivation for something to exist in nature, so we have
emphasized the evidence that actually depends on data.

\section{Current limits on superpartner masses}

The general limits from direct experiments that could produce
superpartners are not very strong.  They are also all model
dependent, sometimes a little and sometimes very much.  Limits from
LEP on charged superpartners are near the kinematic limits except for
certain models, unless there is close degeneracy of the charged
sparticle and the LSP, in which case the decay products are very soft
and hard to observe, giving weaker limits.  So in most cases
charginos and charged sleptons have limits of about 95 GeV.  Gluinos
and squarks have typical limits of about 250 GeV, except that if one
or two squarks are lighter the limits on them are much weaker.  For
stops and sbottoms the limits are about 85 GeV separately. 

There are no general limits on neutralinos, though sometimes such limits
are quoted.  It is clear no general limits exist --- suppose the LSP was
pure photino.  Then it could not be produced at LEP through a Z which
does not couple to photinos, and suppose selectrons were very heavy so
it's production via selectron exchange is very small in pair or
associated production.  Then no cross section at LEP is large enough to
set limits.  There are no general relations between neutralino masses
and chargino or gluino masses, so limits on the latter do not imply
limits on neutralinos.  In typical models the limits are $M_{LSP}\gtrsim
40$ GeV, $M_{\widetilde{N}_{2}}\gtrsim 85$ GeV.  Superpartners get mass
from both the Higgs mechanism and from supersymmetry breaking, so one
would expect them to typically be heavier than SM particles.  All SM
particles would be massless without the Higgs mechanism, but
superpartners would not.  Many of the quark and lepton masses are small
presumably because they do not get mass from Yukawa couplings of order
unity in the superpotential, so one would expect naively that the normal
mass scale for the Higgs mechanism was of order the Z or top masses.  In
models chargino and neutralino masses are often of order Z and top
masses, with the colored gluino mass a few times the Z mass.

There are no firm indirect limits on superpartner masses.  If the
$g_{\mu}-2$ deviation from the SM persists as the data and theory
improve the first such upper limits will be deduced.  If in fact
supersymmetry explains all that we argue above it is explaining,
particularly the EWSB, then there are rather light upper limits on
superpartner masses, but they are not easily made precise. 
Basically, what is happening is that EWSB produces the Z mass in terms
of soft-breaking masses, so if the soft-breaking masses are too large
such an explanation does not make sense.  The soft parameters that
are most sensitive to this issue are $M_{3}$ (basically the gluino
mass) and $\mu$ which strongly affects the chargino and neutralino
masses.  Qualitatively one therefore expects rather light gluino,
chargino, and neutralino masses.  If one takes this argument
\textit{seriously} one is led to expect $M_{\tilde{g} }\lesssim 500$
GeV; $M_{\widetilde{N}_{2}},$ $M_{\widetilde{C}}\lesssim 250$ GeV; and
$M_{\widetilde{N}_{1}}\lesssim 100$ GeV.  These are upper limits,
seldom saturated in models.  There are no associated limits on
sfermions.  They suggest that these gaugino states should be produced in
significant quantities at the Tevatron in the next few years. 

There are some other clues that some superpartners may be light.  If
the baryon number is generated at the EW phase transition then the
lighter stop and charginos should be lighter than the top.  If the
LSP is indeed the cold dark matter, then at least one scalar fermion
is probably light enough to allow enough annihilation of relic LSPs,
but there are loopholes to this argument.

\section{What can supersymmetry explain?}
Supersymmetry can explain much that the SM cannot, as described above,
particularly the Higgs physics as we will discuss in detail below.
Sometimes people who do not understand supersymmetry say it can
``explain or fit anything''.  In fact it is the opposite.
Supersymmetry is a full theory, and all that is unknown is the masses
(which are matrices in flavor space) and the vacuum expectation
values, exactly as for the SM.  There are many conceivable phenomena
that supersymmetry could not explain, including sharp peaks in spectra
at colliders, a world with no Higgs boson below about 200 GeV, a top
quark lighter than the W, deviations from SM predictions greater than
about 1\% for any process with a tree-level SM contribution (including
Z decay to $c\bar{c}$), leptoquarks, wide WW or ZZ resonances, excess
high-P$_{t}$ leptons at HERA, large violations of $\mu/e$
universality, and much more.  None of these has occurred, consistent
with supersymmetry, but a number of them have been reported and then
gone away, and supersymmetry did not ``explain'' them while they were
around.  Supersymmetry alone also cannot explain some real questions
such as why there are three families or the $\mu-\tau$ mass ratio.

\section{How does flavor physics enter the theory?}

The ``flavor problem'' is one of the most basic questions in physics.
 By this usually three questions are intended.  First, why are there
 three families of quarks and leptons, and not more or less?  Second,
 why are the symmetry eigenstates different from the mass eigenstates?
 Third, why do the quarks and leptons have the particular mass values
 they do?  Supersymmetry does not provide the answers to those questions
 directly, though it will affect the answers.  The second and third
 questions are of course related, but different.  We could know the
 answer to the second question but not the third.  For example, the
 actual values of masses of the lighter quarks and leptons could depend
 on operators beyond the tree level in the superpotential.  The u,d,e
 masses are so small that they could get large corrections from a number
 of sources.

Where to look for those answers is not something that is agreed on ---
many people have tried to understand flavor physics at the TeV scale.
Supersymmetry does suggest where to find the answers.  Supersymmetry is
like the SM in that it accommodates the three families and the flavor
rotations but does not explain them.  It clearly suggests that the
flavor physics has basically entered once the superpotential is
determined, i.e. when the Yukawa couplings in the superpotential are
fixed.  That occurs as soon as a 4D theory is written and depends in a
basic way on the string physics and on the compactification and on the
determination of the string vacuum.  Since the superpotential of the
observable sector does not know about supersymmetry breaking, the basic
flavor physics probably does not depend on supersymmetry breaking
either, though how the flavor physics manifests itself in $L_{soft}$
may.  That in turn suggests that learning the Yukawa couplings and the
off-diagonal structure of the trilinears and squark and slepton mass
matrices can guide us to the formulation of how to compactify and how to
find the string vacuum, and can test ideas about such physics.  The role
of supersymmetry breaking is unclear.  For example, the structure of the
trilinear soft-breaking terms can be calculated in terms of the Yukawa
couplings and their derivatives, but may depend on how supersymmetry is
broken as well.

An important point is that we are likely to learn more from data on
the superpartner masses than we did from the quark masses (as we will
discuss later).  That is because the parameters of $L_{soft}$ are rather
directly related to an underlying theory, while the quark and lepton
masses probably are not.  Probably what we learned from the fermion
masses is that some Yukawa couplings are of order unity while others are
small at tree level, arising from non-renormalizable operators and/or
breaking of discrete symmetries and supersymmetry.  The masses of the
first and second family quarks and leptons are probably determined by or
very sensitive to small effects that are hard to calculate (the first
family masses are in the MeV range, while the theory makes sense for the
100 GeV range), while the squark and slepton masses, and probably the
phases, and the approximate size of off-diagonal flavor dependent squark
and slepton masses and trilinears all generally emerge from the theory
at leading order and are thus much more easily interpretable than the
fermion masses.

The next question is how to measure the flavor-dependent elements of
$L_{soft},$ which has 112 flavor-dependent parameters not counting
neutrino physics.  Although certain combinations of them affect
collider physics, and the masses of the mass eigenstates can be
measured at colliders, most of them affect rare decays, mixing, and CP
violation experiments.  Collider studies of superpartners may tell us
little about flavor physics directly.  If they are to have an
observable effect, of course, the supersymmetric contributions to the
decays and mixing and CP violation must be significant, which is most
likely for processes that are forbidden at tree level such as
$b\rightarrow s+\gamma,$ mixing, penguin diagrams, $\mu\rightarrow
e+\gamma,$ etc. 

The absence of flavor-changing decays for many systems puts strong
constraints on some soft parameters.  If the off-diagonal elements of
the squark or slepton mass matrices and trilinears were of order the
typical squark or slepton masses then in general there would be large
flavor mixing effects, since the rotations that diagonalize the quarks
and charged leptons need not diagonalize the squarks and sleptons.
However, many of the constraints from flavor-changing processes in the
literature have been evaluated with assumptions that may not apply, so
people should reevaluate them for any approach they find attractive for
other reasons.  Much effort has gone into constructing models of
$L_{soft}$ that guarantee without tuning the absence of FCNC, and
several approaches exist.  If one of them is confirmed when data exists
it will be a major clue to the structure of the high energy theory.  Our
view that the flavor physics is determined at the high scale implies
that the resulting structure of the squark and slepton mass matrices,
and the trilinear coefficients, is also determined at the high scale and
not by TeV-scale dynamics.  Thus the absence of FCNC is not and should
not be explained by an effective supersymmetric theory.  Rather, the
pattern of soft-breaking terms that is measured and gives small FCNC
will help us learn about the underlying (presumably string) theory.
Similar remarks could be made about proton decay.

Once the soft flavor parameters are measured it is necessary to deduce
their values at the unification or string scales in order to compare
with the predictions of string-based models, or to stimulate the
development of string-based models.  There are two main issues that
arise.  One is how to relate measured values of the CKM matrix and soft
parameters to the values of Yukawa matrices and soft parameters at the
unification scale, assuming no other physics enters between the scales.
This is subtle because the number of independent parameters is
considerably less than the number of apparent parameters in $L_{soft}$
and the superpotential Yukawas, as discussed in Section 17, and the RGE
running will for a generic procedure involve non-physical parameters.
This problem has recently been solved \cite{5}, giving a practical
technique to convert measurements into the form of the high scale
theory.

The second issue is that presumably there is not a desert between the
high and low scales.  Both gauge coupling unification and radiative
electroweak symmetry breaking imply that no part of the theory becomes
strongly interacting below the unification scale.  But we expect
heavy RH neutrinos, axion physics, and perhaps ``exotic'' states such
as those often generated in stringy models, e.g. vector multiplets,
fractionally charged uncolored fermions, etc.  This issue has not
been studied much \cite{6}.  Perhaps by examining appropriate combinations
of quantities for the RGE running, and by imposing appropriate
conditions, it will be possible to use consistency checks to control
the effects of intermediate scale physics.

\section{Derivation of the supersymmetry Lagrangian}

In order to understand the predictions and explanations of
supersymmetry, particularly for the Higgs sector, we must learn the
derivation of the supersymmetry Lagrangian.  I will present the
arguments fully though not all the algebra. I will largely follow the
approach of Martin.

Consider a massless and therefore two-component fermion, $\psi$ whose
superpartner is a complex scalar $\phi.$  Both have two real degrees of
freedom.  But in the off-shell field theory the fermion is a four-component
field with four degrees of freedom, and we want supersymmetry to hold for the
full field theory.  So we introduce an additional complex scalar $F$ so that
there are four scalar degrees of freedom also.  $F$ is called an auxiliary
field.  The combined fields $(\psi,\phi,F)$ are called a chiral superfield or
chiral supermultiplet.  I will not be systematic or careful about the
two-component vs. four-component notation since the context usual is
clear.  The Lagrangian can be taken to be

\begin{eqnarray}
-L_{free}=\sum_{i}(\partial^{\mu}\phi_{i}^{\ast}\partial_{\mu}\phi_{i}
+\bar{\psi}_{i}\gamma^{\mu}\partial_{\mu}\psi_{i}+F_{i}^{\ast}F_{i}).
\end{eqnarray}

The sum is over all chiral supermulitplets in the theory.  Note that
the dimensions of $F$ are $[F]=m^{2}.$  The Euler-Lagrange equations
of motion for $F$ are $F=F^{\ast}=0,$ so on-shell we revert to only
two independent degrees of freedom.  One can define supersymmetry
transformations that take bosonic degrees of freedom into fermionic
ones; we will look briefly at them later.  The supersymmetry
transformations can be defined so that $L_{free}$ is invariant.  Next
we write the most general set of renormalizable interactions,

\begin{eqnarray}
L_{chiral}=L_{free}+L_{int}
\end{eqnarray}

\begin{eqnarray}
L_{int}=-\frac{1}{2}W^{ij}\psi_{i}\psi_{j}+W^{i}F_{i}+c.c.
\end{eqnarray}

Here $W^{ij}$ and $W^{i}$ are any functions of only the scalar fields,
remarkably, and $W^{ij}$ is symmetric.  If $W^{ij}$ or $W^{i}$
depended on the fermion or auxiliary fields the associated terms would
have dimension greater than four, and would therefore not be
renormalizable.  There can be no terms in $L_{int}$ that depend on
$\phi_{i}^{\ast}$ or $\phi_{i}$ since such terms would not transform
into themselves under the supersymmetry transformations.

Now imagine supersymmetry transformations that mix fermions and
bosons,
$\phi\rightarrow\phi+\varepsilon\psi,\psi\rightarrow\psi+\varepsilon\phi$.
 We should go through these transformations in detail with indices,
but one can see the basic argument simply.  Here $\varepsilon$ must
be a spinor so each term behaves the same way in spin space, and we
can take $\varepsilon$  to be a constant spinor in space-time, and
infinitesimal.  Then the variation of the Lagrangian (which must
vanish or change only by a total derivative if the theory is invariant
under the supersymmetry transformation) contains two terms with four
spinors:

\begin{eqnarray} 
\delta L_{int}=-\frac{1}{2}\frac{\delta W^{ij}}{\delta\phi_{k}}(\varepsilon
\psi_{k})\psi_{i}\psi_{j}-\frac{1}{2}\frac{\delta
W^{ij}}{\delta\phi_{k}
^{\ast}}(\varepsilon^{\dagger}\psi_{k}^{\dagger})\psi_{i}\psi_{j}
+c.c.
\end{eqnarray}

Neither term can cancel against some other term.  For the first term
there is a Fierz identity
$(\varepsilon\psi_{i})(\psi_{j}\psi_{k})+(\varepsilon\psi
_{j})(\psi_{k}\psi_{i})+(\varepsilon\psi_{k})(\psi_{i}\psi_{j})=0$, so
if and only if $\delta W^{ij}/\delta\phi_{k}$ is totally symmetric
under interchange of i,j,k the first term vanishes identically \ For
the second term the presence of the hermitean conjugation allows no
similar identity, so it must vanish explicitly, which implies $\delta
W^{ij}/\delta\phi_{k}^{\ast}=0,$and thus $W^{ij}$ cannot depend on
$\phi^{\ast}$!  $W^{ij}$ must be an analytic function of the complex
field $\phi.$

Therefore we can write

\begin{eqnarray}
W^{ij}=M^{ij}+y^{ijk}\phi_{k},
\end{eqnarray}

\noindent where $M^{ij}$ is a symmetric matrix that will be the
fermion mass matrix, and $y^{ijk}$ can be called Yukawa couplings
since it gives the strength of the coupling of boson $k$ with fermions
$i,j$; $y^{ijk}$ must be totally symmetric.  Then it is very
convenient to define

\begin{eqnarray}
W=\frac{1}{2}M^{ij}\phi_{i}\phi_{j}+\frac{1}{6}y^{ijk}\phi_{i}\phi_{j}\phi
_{k}
\end{eqnarray}

\noindent and $W^{ij}=\delta^{2}W/\delta\phi_{i}\delta\phi_{j}.$ $W$
is the ``superpotential'', an analytic function of $\phi$, and a
central function of the formulation of the theory.  $W$ is fully
supersymmetric and gauge invariant and Lorentz invariant, and an
analytic function of $\phi$ (i.e. it cannot depend explicitly on
$\phi^{\ast}$), so it is highly constrained.  It determines the most
general non-gauge interactions of the chiral superfields.

A similar argument for the parts of $\delta L_{int}$ which contain a
spacetime derivative imply that $W^{i}$ is determined in terms of $W$
as well,

\begin{eqnarray}
W^{i}=\frac{\delta W}{\delta\phi_{i}}=M^{ij}\phi_{j}+\frac{1}{2}y^{ijk}
\phi_{j}\phi_{k}.
\end{eqnarray}

\noindent Because interactions are now present, the equations for $F$
are non-trivial,

\begin{eqnarray}
F_{i}=-W_{i}^{\ast}.
\end{eqnarray}

\noindent The scalar potential is related to the Lagrangian by $L=T-V,$ so

\begin{eqnarray}
V=\sum_{i}\left|  F_{i}\right|  ^{2}
\end{eqnarray}

\noindent This contribution is called an ``F-term'', and is
automatically bounded from below, an important improvement.

The above analysis was appropriate for chiral superfields, which will contain
the fermions and their superpartners.  Now we repeat the logic for the gauge
supermultiplets that contain the gauge bosons and their superpartners.
 Initially they are massless gauge bosons, like photons, $A_{\mu}^{a},$ with
gauge index $a,$ and two degrees of freedom.  Their superpartners are
two-component spinors $\lambda^{a}.$  But as above, off shell the fermion has
four degrees of freedom, while the massive boson has three, the two transverse
polarizations and a longitudinal polarization.  So again it is necessary to
add an auxiliary field, a real one since only one degree of freedom is needed,
called $D^{a}.$  Then the Lagrangian has additional pieces

\begin{eqnarray}
L_{gauge}=-\frac{1}{4}F_{\mu\nu}^{a}F_{a}^{\mu\nu}-i\lambda^{\dagger a}
\gamma^{\mu}D_{\mu}\lambda^{a}+\frac{1}{2}D^{a}D^{a},
\end{eqnarray}

\noindent where

\begin{eqnarray}
F_{\mu\nu}^{a}=\partial_{\mu}A_{\nu}^{a}-\partial_{\nu}A_{\mu}^{a}
-gf^{abc}A_{\mu}^{b}A_{\nu}^{c}
\end{eqnarray}

\noindent and the covariant derivative is

\begin{eqnarray}
D_{\mu}\lambda^{a}=\partial_{\mu}\lambda^{a}-gf^{abc}A_{\mu}^{b}\lambda
^{c}.
\end{eqnarray}

Note that the notation is unfortunate, with both the covariant
derivative and the new field being denoted by the standard ``$D$''.
Also, I have not been careful about two component vs. four component
spinors.  It is crucial for gauge invariance that the same coupling $g$
appears in the definition of the tensor $F$ and in the covariant
derivative.  Lagrangians always have to contain all of the terms allowed by
gauge invariance, etc., and here we can see another term to add,

\begin{eqnarray} 
(\phi_{i}^{\ast}T^{a}\phi_{i})D^{a}.
\end{eqnarray}

\noindent There is one more term that can be added that mixes the fields,
$\lambda ^{\dagger a}(\psi^{\dagger}T^{a}\phi),$ and its conjugate,
with an arbitrary dimensionless coefficient.  Requiring the entire
Lagrangian to be invariant under supersymmetry transformations
determines the arbitrary coefficient and gives a resulting Lagrangian

\begin{eqnarray}
L=L_{gauge}+L_{chiral}+g_{a}(\phi^{\ast}T^{a}\phi)D^{a}-\sqrt{2}g_{a}
[(\phi^{\ast}T^{a}\psi)\lambda^{a}+\lambda^{\dagger a}(\psi^{\dagger}T^{a}
\phi)]
\end{eqnarray}

\noindent where all derivatives in earlier forms are replaced by
covariant ones. Remarkably, the requirement of supersymmetry fixed the
couplings of the last terms to be gauge couplings even though they are
not normal gauge interactions!  The equations of motion for $D^{a}$
give $D^{a}=-g(\phi^{\ast }T^{a}\phi),$ so the scalar potential is

\begin{eqnarray}
V=F^{\ast i}F_{i}+\frac{1}{2}\sum_{a}D^{a}D^{a}=\left|  \partial
W/\partial\phi_{i}\right|  ^{2}+\sum_{a}g_{a}^{2}(\phi^{\ast}T^{a}\phi
)^{2}.
\end{eqnarray}

\noindent The sum is over $a=1,2,3$ for the three gauge couplings.  The
two terms are called ``F-terms'' and ``D-terms''.  Remarkable, the
unbroken supersymmetric theory gives a scalar potential bounded from
below.  On the one hand that is good since unbounded potentials are a
problem, but it also implies that the Higgs mechanism cannot happen for
unbroken supersymmetry since the potential will be minimized at the
origin.  In the above,

\begin{eqnarray}
L_{chiral}=D^{\mu}\phi_{i}^{\ast}D_{\mu}\phi_{i}+\bar{\psi}_{i}\gamma^{\mu
}D_{\mu}\psi_{i}
\end{eqnarray}

\begin{eqnarray*}
+(\frac{1}{2}M_{ij}\psi_{i}\psi_{j}+\frac{1}{2}y^{ijk}\phi
_{i}\psi_{j}\psi_{k}+c.c.)+F_{i}^{\ast}F_{i}.
\end{eqnarray*}

\noindent This completes the derivation of the unbroken supersymmetry
Lagrangian.

\section{Non-renormalization theorem}

For unbroken supersymmetry there is a very important result, called the
non-renormalization theorem, that is very useful for building models to
relate the theory to the real world.  Because of this result the
supersymmetry fields get a wave function renormalization only, so they
have the familiar log running of couplings and masses, but no other
renormalizations.  Consequently the parameters of the superpotential $W$
are not renormalized, in any order of perturbation theory.  In
particular, terms that were allowed in $W$ by gauge invariance and
Lorentz invariance are not generated by quantum corrections if they are
not present at tree level$,$ so no F-terms are generated if they are
initially absent.  If there is no $\mu$ --term in the superpotential
(see below), none is generated.  The non-renormalization theorem is
difficult to probe without extensive formalism, so I just state it here.
References and a pedagogical derivation are given in reference 7.

\section{Toward softly-broken supersymmetry with a toy model}

Consider the Wess-Zumino model, with,

\begin{eqnarray}
W=\frac{m}{2}\phi \phi +\frac{g}{6}\phi \phi \phi ,
\end{eqnarray}

\noindent and

\begin{eqnarray}
L=(\partial \phi )^{2}+i\Psi ^{\dagger }\bar{\sigma}^{\mu }\partial _{\mu
}\Psi -F_{\phi }^{\ast }F_{\phi }+(\frac{1}{2}W_{\phi \phi }\Psi \Psi
-W_{\phi }F_{\phi }+c.c.).
\end{eqnarray}

This is written in two component notation. $W_{\phi }=-F_{\phi }^{\ast
}=m\phi +\frac{g}{2}\phi \phi $ is the derivative of the superpotential
with respect to $\phi ,$and $W_{\phi \phi }$ the second derivative.  We
put $ \phi =(A+iB)/2$ and $F_{\phi }=(F+iG)/2,$ where $A,B,F,G$ are real
scalars, and switch to four component notation.  Under the supersymmetry
transformations, with $\varepsilon $ a constant spinor,

\begin{eqnarray}
%\begin{align}
\delta A& =\bar{\varepsilon}\gamma _{5}\Psi ,  \\
\delta B& =-\bar{\varepsilon}\Psi , \\
\delta \Psi & =F\varepsilon -G\gamma _{5}\varepsilon +\gamma ^{\mu }\partial
_{\mu }\gamma _{5}A\varepsilon +\gamma ^{\mu }\partial _{\mu }B\varepsilon ,
\\
\delta F& =-\bar{\varepsilon}\gamma ^{\mu }\partial _{\mu }\Psi , \\
\delta G& =-\bar{\varepsilon}\gamma _{5}\gamma ^{\mu }\partial _{\mu }\Psi ,
%\end{align}
\end{eqnarray}

\noindent the Lagrangian changes by a total derivative, so the action is
invariant with the usual assumptions.   

Now substitute for $W_{\phi }$ and $W_{\phi \phi }$ etc.  Then the
Lagrangian is

\begin{eqnarray*}
L =\frac{1}{2}(\partial A)^{2}+\frac{1}{2}(\partial B)^{2}+\frac{i}{2}\bar{
\psi}\gamma ^{\mu }\partial _{\mu }\psi +\frac{1}{2}m\bar{\psi}\psi  
\end{eqnarray*}
\begin{eqnarray*}
+\frac{g}{\sqrt{2}}A\bar{\psi}\psi -\frac{ig}{\sqrt{2}}B\bar{\psi}\gamma
_{5}\psi-\frac{1}{2}(F^{2}+G^{2})
\end{eqnarray*}
\begin{eqnarray}
\hspace{.7in}-\frac{m}{2}(2AF-2BG)-\frac{g}{2\sqrt{2}}(F(A^{2}-B^{2})-2GAB).
\end{eqnarray}

\noindent Now the equations of motion for $F,G$ are

\begin{eqnarray}
F=-mA-\frac{g}{2\sqrt{2}}(A^{2}-B^{2}),G=mB+\frac{g}{\sqrt{2}}AB.
\end{eqnarray}

\noindent Substituting these gives interaction vertices $\frac{mg}{2\sqrt{2}}
A(A^{2}-B^{2}).$

With one coupling strength $g$ and one mass $m$ the full Lagrangian is
supersymmetric.  (Note that without supersymmetry there can be four
different masses and four different couplings, so there are six relations
predicted by supersymmetry which only allows one mass and one coupling.) 
But when supersymmetry is broken we expect the masses to differ.  Suppose
we allow four different masses, $m_{A},m_{B},m_{\psi },$ and $m_{g},$ where
the last is the mass that is needed in some terms to give each term
dimension four, so it multiplies $g.$  It's clear how to rewrite the
Lagrangian with these separate masses.  There are four three-particle
vertices, $A\bar{\psi}\psi $, $A^{3},AB^{2},B\bar{\psi}\psi .$  Now if we
write the expression for a tadpole graph,

\begin{eqnarray*}
\hspace{-3.8in}\langle 0\left| L\right| A\rangle =
\end{eqnarray*}
\begin{eqnarray}
\hspace{.4in}\frac{g}{\sqrt{2}}\left\{ 4m_{\psi }\int 
\frac{d^{4}p}{p^{2}-m_{\psi }^{2}}-m_{g}\int \frac{d^{4}p}{p^{2}-m_{B}^{2}}
-3m_{g}\int \frac{d^{4}p}{p^{2}-m_{A}^{2}}\right\},
\end{eqnarray}

\noindent we see that in general this has a quadratic divergence, which
cancels in the supersymmetry limit as expected.  The fermion loop gives
a minus sign, the factor of 4 in the first term arises from $Tr(\gamma
^{\mu }p_{\mu }+m_{\psi })=Trm_{\psi }=4m_{\psi }$, and the 3 in the
last from the $A^{3}.$ But --- and here is the important point --- the
divergence still cancels if $m_{A}\neq m_{B}\neq m_{g},$ but not if
$m_{\psi }\neq m_{g}.$ Thus extra contributions to boson masses do not
reintroduce quadratic divergences --- they are called ``soft''
supersymmetry breaking.  But extra contributions to fermion masses do
lead to quadratic divergences, ``hard'' supersymmetry breaking.  This
result is true to all orders in perturbation theory, though this
pedagogical argument does not show it.  Some of the results are obvious
since couplings proportional to masses will not introduce quadratic
divergences, but it is still helpful to see the supersymmetry
structure.  After the supersymmetry breaking there are three masses and
one coupling, so there are still four tests that the theory is a broken
supersymmetric one.

To understand the general structure of supersymmetry breaking better, recall
how symmetry breaking works in the SM.  It is not possible to break the $
SU(2)\times U(1)$ symmetry from within the SM.  So a new sector, the Higgs
sector is needed.  Interactions are assumed in the Higgs sector that give a
potential with a minimum away from the origin, so the Higgs field gets a vev
which breaks the symmetry.  To generate mass for $W,Z,q,l$ an interaction
is needed to transmit the breaking to the ``visible'' particles $W,Z,q,l.$ 
For fermions this interaction is $L_{fermion}=g_{e}\bar{e}
_{L}e_{R}h+cc\rightarrow g_{e}$v$\bar{e}e$ after h gets a vev for the
fermions, and we can identify $m_{e}=g_{e}$v.  Similarly, for the gauge
bosons the Lagrangian term $(D^{\mu}h)(D_{\mu}h)\rightarrow
g^{2}hhW^{\mu}W_{\mu}\rightarrow g^{2}$v$^{2}W^{\mu}W_{\mu}$ giving $W,Z$
masses.  The fundamental symmetry breaking is spontaneous (h gets a vev),
but the effective Lagrangian appears to have explicit breaking.

The situation is very similar for supersymmetry.  It is not possible to
break supersymmetry in the ``visible'' sector, i.e. the sector containing
the superpartners of the SM particles.  A separate sector is needed where
supersymmetry is broken.  Originally it was called the ``hidden'' sector,
but that is not a good name since it need not be really hidden.  Then there
must be some interaction(s) to transmit the breaking to the visible sector.
 Since the particles of both sectors interact gravitationally, gravity can
always transmit the breaking. Other interactions may as well.  We will
have to find out how the breaking is transmitted from data on the
superpartners, their masses and decays and phases and flavor rotations. 
Different ways of transmitting the breaking give different patterns of the
soft parameters that we discuss below.  A significant complication is that
the effects of the supersymmetry breaking are mixed up with effects of the
transmission.  All the effects of the supersymmetry breaking and of the way
it is transmitted, for any theory, are embedded in the soft-breaking
Lagrangian that we turn to studying.

\section{The soft-breaking Lagrangian}

The (essentially) general form of $L_{soft}$ is \cite{8}

\begin{eqnarray}
L_{soft}=\frac{1}{2}(M_{\lambda}\lambda^{a}\lambda^{a}+c.c.)+m_{ij}^{2}
\phi_{j}^{\ast}\phi_{i}
\end{eqnarray}
\begin{eqnarray*}
+(\frac{1}{2}b_{ij}\phi_{i}\phi_{j}+\frac{1}{6}
a_{ijk}\phi_{i}\phi_{j}\phi_{k}+c.c.)
\end{eqnarray*}

\noindent This obviously breaks supersymmetry since only scalars and
gauginos get mass, not their superpartners.  It is soft as in our
example above because it can be proved to not introduce any quadratic
divergences.  Models for supersymmetry breaking, however they
originate, in string theory or supergravity or dynamically, all lead
to this form.  We will write it for the SSM shortly.

If all fields carry gauge quantum numbers there are terms that could
be added to this without generating quadratic divergences, such as
$\phi_{i}^{\ast} \phi_{j}\phi_{k},$ but such terms seldom arise in
models so they are usually ignored \cite{9}.  If such terms are truly
absent once measurements are analyzed, their absence may be a clue to
how supersymmetry is broken and transmitted.

\section{{\Large T}he {\Large M}inimal {\Large S}upersymmetric {\Large
S}tandard {\Large M}odel}

To write the supersymmetric SM we first take all of the quarks and
leptons and put them in chiral superfields with superpartners.  For
each set of quantum numbers, such as up quarks or electrons, the
scalar, fermion, and auxiliary fields $(\phi,\psi,F)$ form a
supermultiplet in the same sense as $(n,p)$ form a strong isospin
doublet or $(\nu_{e},e)$ form  an electroweak doublet.  All
superpartners are denoted with a tilde, and there is a superpartner
for each spin state of each fermion --- that is important since the SM
treats fermions of different chirality differently.  The gauge bosons
are put in vector superfields with their fermionic superpartners. 
Since $W$ is analytic in the scalar fields, we cannot include the
complex conjugate of the scalar field as in the SM to give mass to the
down quarks, so there must be two Higgs doublets (or more) in
supersymmetry, and each has its superpartners.  The requirement that
the trace anomalies vanish so that the theories stay renormalizable,
$TR(Y^{3})=TR(T_{3L}^{2}Y)=0,$ also implies the existence of the same
two Higgs doublets. (The relevance of anomalies may seem unclear since
we are only writing an effective theory, while anomaly conditions only
need to be satisfied for the full theory.  But if the anomaly conditions
are not satisfied it may introduce a sensitivity to higher scales that
the effective theory should not have.)

We proceed by first constructing the superpotential so we can
calculate the F-terms, and then writing the Lagrangian, following
equation 6 and summing over all the particles.  The most general
superpotential, if we don't extend the SM and don't include RH
neutrinos, is 

\begin{eqnarray}
W=\bar{u}Y_{u}QH_{u}-\bar{d}Y_{d}QH_{d}-\bar{e}Y_{e}LH_{d}+\mu H_{u}
H_{d}.
\end{eqnarray}

\noindent All the fields are chiral superfields.  The bars over
$u,d,e$ are in the sense of Martin's notation, specifying the
conjugate fields.  The signs are conventional so that masses later are
positive.  Indices are suppressed --- for example, the fourth and
first terms are

\begin{eqnarray}
\mu(H_{u})_{\alpha}(H_{d})_{\beta}\varepsilon_{\alpha\beta}\;{\rm and}\;\bar
{u}_{ai}(Y_{u})_{ij}Q_{j\alpha}^{a}(H_{u})_{\beta}\varepsilon_{\alpha\beta
}.
\end{eqnarray}

The Yukawa couplings $Y_{u}$ etc. are dimensionless 3$\times3$ family
matrices that determine the masses of quarks and leptons, and the
angles and phase of the CKM matrix after $H_{u}^{0}$ and $H_{d}^{0}$
get vevs.  They also contribute to the squark-quark-higgsino
couplings etc. since the fields in $W$ are superfields containing all
the components.  This is the most general superpotential for the SSM
if we assume baryon and lepton number are conserved (we'll return to
this question).  To see the structure more explicitly we can use the
approximations

\begin{eqnarray}
Y_{u}\approx\left(
\begin{array}
[c]{ccc}
0 & 0 & 0\\
0 & 0 & 0\\
0 & 0 & Y_{t}
\end{array}
\right)  ,{ \ \ \ \ }Y_{d}\approx\left(
\begin{array}
[c]{ccc}
0 & 0 & 0\\
0 & 0 & 0\\
0 & 0 & Y_{b}
\end{array}
\right)  ,{ \ \ \ \ }Y_{e}\approx\left(
\begin{array}
[c]{ccc}
0 & 0 & 0\\
0 & 0 & 0\\
0 & 0 & Y_{\tau}
\end{array}
\right)  ,
\end{eqnarray}

\noindent which gives

\begin{eqnarray}
W=Y_{t}(\bar{t}tH_{u}^{0}-\bar{t}bH_{u}^{+})-Y_{b}(\bar{b}tH_{d}^{-}-\bar
{b}bH_{d}^{0})
\end{eqnarray}
\begin{eqnarray*}
\hspace{.5in}-Y_{\tau}(\bar{\tau}\nu_{\tau}H_{d}^{-}-\bar{\tau}\tau H_{d}
^{0})+\mu(H_{u}^{+}H_{d}^{-}-H_{u}^{0}H_{d}^{0})
\end{eqnarray*}

There are also other interactions from $W$ such as vertices
$H_{u}^{0} t_{R}^{\ast}t_{L},$
$\tilde{H}_{u}^{0}t_{R}^{\ast}\tilde{t}_{L},$ $\tilde
{H}_{u}^{0}\tilde{t}_{R}t_{L},$ etc., all with the same strength
$Y_{t}.$  All of them are measurable, and it will be an important
check of supersymmetry to confirm they are all present with the same
strength.  All are dimensionless, so supersymmetry-breaking will only
lead to small radiative corrections to these coupling strengths.  In
general one goes from one to another of these by changing any pair of
particles into superpartners.

Before we turn to writing the full soft-breaking Lagrangian, we first look at
two significant issues that depend on how supersymmetry is embedded in a more
basic theory. 

\section{The $\mu$ opportunity}

The term $\mu H_{u}H_{d}$ in the superpotential leads to a term in the
Lagrangian

\begin{eqnarray}
L=........+\mu(\tilde{H}_{u}^{+}\tilde{H}_{d}^{-}-\tilde{H}_{u}^{0}\tilde
{H}_{d}^{0})+.......
\end{eqnarray}

\noindent which gives mass terms for higgsinos in the chargino and
neutralino mass matrices, so $\mu$ enters there.  This term also
contributes to the scalar Higgs potential from the F-terms,

\begin{eqnarray}
V=.......\left|  \mu\right|  ^{2}(\left|  H_{u}^{0}\right|  ^{2}+\left|
H_{d}^{0}\right|  ^{2}+...)+........
\end{eqnarray}

\noindent so these terms affect the Higgs mass, and F-terms also give
contributions to the Lagrangian that affect the squark and slepton
mass matrices,

\begin{eqnarray}
L=.....\mu^{\ast}(\widetilde{\bar{u}}Y_{u}\tilde{u}H_{d}^{0\ast}
+....)+....
\end{eqnarray}

Thus phenomenologically $\mu$ must be of order the weak scale to
maintain the solutions of the hierarchy problem, gauge coupling
unification, and radiative electroweak symmetry breaking.  The naive
scale for any term in the superpotential is one above where the
supersymmetry is broken, e.g. the string scale or unification scale,
and since $\mu$ occurs in $W$ one would naively expect $\mu$ to be of
order that scale, far above the weak scale.  In the past that has
been called the ``$\mu$ problem''.  But actually it is a clue to the
correct theory and is an opportunity to learn what form the underlying
theory must take.  For example, in a string theory we expect all the
mass terms to vanish since the SM particles are the massless modes of
the theory, so in a string theory $\mu$, which is a mass term, would
naturally vanish.  That could be a clue that the underlying theory is
indeed a string theory.  In the following we will view $\mu=0$ as a
``string boundary condition''.  Older approaches added symmetries to
require $\mu=0.$  Note that because of the non-renormalization
theorem once $\mu$ is set to zero in $W$ it is not generated by loop
corrections. 

We also know phenomenologically that the $\mu$ contribution to the
chargino and neutralino masses and the Higgs mass cannot vanish, or
some of them would be so light they would have been observed, so we
know that somehow a piece that plays the same role as $\mu$ is
generated.  We will call it $\mu_{eff},$ but whenever there is no
misunderstanding possible we will drop the subscript and just write
$\mu$ for $\mu_{eff}.$  Different ways of generating $\mu_{eff}$ give
different relations to the other soft-breaking parameters, a different
phase for $\mu_{eff},$ a characteristic size for $\mu_{eff},$ etc.  
Once it is measured we will have more clues to the underlying
theory.  Any top-down approach must generate $\mu_{eff}$ and its
phase correctly.

\section{R-parity conservation}

The $\mu$ opportunity looks like the $\mu$ problem if one views
supersymmetry as an effective low energy theory without seeing it as
embedded in a more fundamental high scale theory.  Similarly, if we
view supersymmetry as only a low energy effective theory there is
another complication that arises.  There are additional terms that
one could write in $W$ that are analytic, gauge invariant, and Lorentz
invariant, but violate baryon and/or lepton number conservation.  No
such terms are allowed in the SM, which accidentally conserves B and L
to all orders in perturbation theory, though it does not conserve them
non-perturbatively.  These terms are

\begin{eqnarray}
W_{R}=\lambda_{ijk}L_{i}L_{j}e_{k}+\lambda_{ijk}^{\prime}L_{i}Q_{j}\bar{d}
_{k}+\lambda_{ijk}^{\prime\prime}\bar{u}_{i}\bar{d}_{j}d_{k}.
\end{eqnarray}

\noindent The couplings
$\lambda,\lambda^{\prime},\lambda^{\prime\prime}$ are matrices in
family space.  Combining the second and third one can get very rapid
proton decay, so one or both of them must be required to be absent.
That is not the way one wants to have a theory behave.  Rather, B and
L conservation consistent with observation should arise naturally from
the symmetries of the theory.  Most, but not all, theorists expect
that an underlying symmetry will be present in the broader case to
forbid all of the terms in $W_{R}.$

There are two approaches to dealing with $W_{R}$.  We can add a
symmetry to the effective low energy theory, called R-parity or a
variation called matter parity, which we assume will arise from a
string theory or extended gauge group.  R-parity is multiplicatively
conserved,

\begin{eqnarray}
R=(-1)^{3(B-L)+2S}
\end{eqnarray}

\noindent where $S$ is the spin.  Then SM particles and Higgs fields
are even, superpartners odd.  This is a discrete Z$_{2}$
symmetry. Such symmetries that treat superpartners differently from SM
particles and therefore do not commute with supersymmetry are called
R-symmetries.  Equivalently, one can use ``matter parity'',

\begin{eqnarray}
P_{m}=(-1)^{3(B-L)}.
\end{eqnarray}

\noindent A term in $W$ is only allowed if $P_{m}=+1.$ Gauge fields
and Higgs are assigned $P_{m}=+1,$ and quark and lepton
supermultiplets $P_{m}=-1.$ $P_{m}$ commutes with supersymmetry and
forbids $W_{R}.$ Matter parity could be an exact symmetry, and such
symmetries do arise in string theory.  If R-parity or matter parity
holds there are major phenomenological consequences,

$\bullet$ At colliders, or in loops, superpartners are produced in pairs.

$\bullet$ Each superpartner decays into one other superpartner (or an
odd number).

$\bullet$ The lightest superpartner (LSP) is stable.  That determines
supersymmetry collider signatures, and makes the LSP a good candidate
for the cold dark matter of the universe.

The second approach is very different, and does not have any of the
above phenomenological consequences.  One arbitrarily sets
$\lambda^{\prime}$ or $\lambda^{\prime\prime}=0$ so there are no
observable violations of baryon number or lepton number
conservation.  Other terms are allowed and one sets limits on them
when their effects are not observed, term by term.  In the MSSM
itself R-parity must be broken explicitly if it is broken at all.  If
it were broken spontaneously by a sneutrino vev there would be a
Goldstone boson associated with the spontaneous breaking of lepton
number (called a Majaron), and some excluded Z decays would have been
observed. 

We will not pursue this ad hoc approach, because we do not like
arbitrarily setting some terms to zero, and we do not like giving up
the LSP as cold dark matter if we are not forced to.  Further, large
classes of theories conserve R-parity or matter parity \cite{10}.  Often
theories have a gauged U(1)$_{B-L}$ symmetry that is broken by scalar
vevs and leaves $P_{m}$ automatically conserved.  String theories
often conserve R-parity or P$_{m}.$  Often theories conserve R-parity
at the minimum of the Higgs potential.  Baryogenisis via
leptogenesis probably requires R-parity conservation because the usual
$B+L$ violation plus $L$ violation would allow the needed asymmetries
to be erased.  The lepton number needed for $\nu$ seesaw masses
violates $L$ by two units and does not violate R-parity
conservation.  In general, when supersymmetry is viewed as embedded
in a more fundamental theory, R-parity conservation is very likely and
easily justified.   Ultimately, of course, experiment will decide,
but we will assume R-parity conservation in the rest of these
lectures.

\section{Definition of MSSM}

At this stage we can define the effective low energy supersymmetry
theory, which we call the MSSM, as the theory with the SM gauge group
and particles, and the superpartners of the SM particles, and
conserved $R$-parity, and two Higgs doublets.  Perhaps it would be
better to include right handed neutrinos and their superpartners as
well, but that is not yet conventional. 

\section{The MSSM soft-breaking Lagrangian}

We can now write the general soft-breaking Lagrangian for the MSSM,

\begin{eqnarray*}
\hspace{-.3in}-L_{soft}  & =\frac{1}{2}(M_{3}\tilde{g}\tilde{g}+M_{2}\widetilde{W}
\widetilde{W}+M_{1}\widetilde{B}\widetilde{B}+c.c.)
\end{eqnarray*}
\begin{eqnarray*}
\hspace{.8in}+\widetilde{Q}^{\dag}m_{Q}^{2}\widetilde{Q}+\widetilde{\bar{u}}^{\dag}m_{\bar{u}}^{2}\widetilde{\bar{u}}+
\widetilde{\bar{d}}^{\dag}m_{d}^{2}\widetilde{\bar{d}}+
\widetilde{L}^{\dag}m_{L}^{2}\widetilde{L}+\widetilde{\bar{e}}^{\dag}m_{\bar{e}}^{2}\widetilde{\bar{e}}
\end{eqnarray*}
\begin{eqnarray*}
\hspace{.5in}+(\widetilde{\bar{u}}^{\dag}a_{u}\widetilde{Q}H_{u}-\widetilde{\bar{d}
}^{\dag}a_{d}\widetilde{Q}H_{d}-\widetilde{\bar{e}}^{\dag}a_{e}\widetilde
{L}H_{d}+c.c.)
\end{eqnarray*}
\begin{eqnarray}
\hspace{.5in}+m_{H_{u}}^{2}H_{u}^{\ast}H_{u}+m_{H_{d}}^{2}H_{d}^{2\ast}+(bH_{u}
H_{d}+c.c.).
\end{eqnarray}

\noindent For clarity a number of the indices are suppressed.
 $M_{1,2,3}$ are the complex bino, wino, and gluino masses, e.g.
$M_{3}=\left| M_{3}\right| e^{i\phi_{3}},$ etc.  In the second line
$m_{Q}^{2},$ etc, are squark and slepton hermitean 3$\times3$ mass
matrices in family space.  The $a_{u,d,e}$ are complex 3$\times3$
family matrices, usually called trilinear couplings.  b is sometimes
written as $B\mu$ or as $m_{3}^{2}$ or as $m_{12}^{2}.$ Additional
parameters come from the gravitino complex mass and from
$\mu_{eff}=\mu e^{i\phi_{\mu}};$ we will usually risk writing the
magnitude of $\mu_{eff}$ as just $\mu$ assuming the context will
distinguish this from the original $\mu$ of the superpotential.
 This may seem to involve a lot of parameters, but all the physical
parameters are observable from direct production and study of
superpartners and their effects.  The absence of observation of
superpartners and their effects already gives us useful information
about some of the parameters.  It is important to understand that all
of these parameters are masses or flavor rotation angles or phases or
Higgs vevs, just as for the SM.  If we had no measurements of the
quark and lepton masses and interactions there would be even more
parameters for the SM than here.

With this Lagrangian we can do general, useful, reliable
phenomenology, as we will see.  For example, in the SM we did not
know the top quark mass until it was measured.  Nevertheless, for any
chosen value of the top mass we could calculate its production cross
section at any collider, all of its decay BR, its contribution to
radiative corrections, etc.  Similarly, for the superpartners we can
calculate expected signals, study any candidate signal and evaluate
whether it is consistent with the theory and with other constraints or
data, and so on.  A possible signal might have too small or large a
cross section to be consistent with any set of parameters, or decay BR
that could not occur here.  Many examples can be given.  We can also
study whether superpartners can be studied at any proposed future
facility.   Further, most processes depend on only a few of the
parameters --- we will see several examples of this in the
following. 

Now let us count the parameters of the broken supersymmetric theory
relative to the SM.  There are no new gauge or Yukawa couplings, and
still only one strong CP angle $\bar{\vartheta},$ so that is already
rather economical.  Then

$\bullet$ $m_{Q}^{2},$ etc are 5 3$\times3$ hermitean matrices $\rightarrow$ 9
real parameters each $\rightarrow$ 45

$\bullet$ $a_{u,d,e}$ are 3 3$\times3$ complex matrices $\rightarrow$ 18 real
parameters each $\rightarrow$ 54

$\bullet$ $M_{1,2,3}$, $\mu,b$ are complex $\rightarrow$ 10

$\bullet$ $m_{H_{u,d}}^{2}$ are real by hermiticity $\rightarrow$ 2

\noindent giving a total of 111 parameters.  As for the CKM quark
matrix it is possible to redefine some fields and absorb some
parameters.  Baryon and lepton number are conserved, and there are two
U(1) symmetries that one can see by looking at the Lagrangian.  One
arises because if $\mu$ and $b$ are zero there is a symmetry where
$H_{u,d}\rightarrow$ $e^{i\alpha}H_{u,d}$ and the combinations
$L\bar{e},Q\bar{u},Q\bar{d}\rightarrow
e^{-i\alpha}L\bar{e},Q\bar{u},Q\bar {d}.$ For example, one can take
$Q\rightarrow e^{-i\alpha}Q,$ $L\rightarrow e^{-i\alpha}L,$ and
$\bar{e},\bar{u},\bar{d}$ invariant.  Such a symmetry is called a
Peccei-Quinn symmetry if it holds for $\mu=0$ but is broken when
$\mu\neq0.$ The other arises because if $M_{i},a_{i},b=0$ there is a
continuous R-symmetry, e.g. the Higgs fields can have charge 2, the
other matter fields charge 0, and the superpotential charge 2.
Symmetries are called R-symmetries whenever members of a
supermultiplet are treated differently.

With these four symmetries, four parameters can be absorbed.  Also,
the SM has two parameters in the Higgs potential,
$\mu^{2}\phi^{2}+\lambda\phi^{4},$ so to count the number beyond the
SM we subtract those 2.  Then there are 111-4-2=105 new parameters.
The SM itself has 3 gauge couplings, 9 quark and charged lepton
masses, 4 CKM angles, 2 Higgs potential parameters, and one strong CP
phase $\rightarrow$ 19.  So there are 124 parameters altogether.
 When massive neutrinos are included one has RH $\nu$ masses, and the angles
of the flavor rotation matrix (which has 3 real angles and 3 phases for the
$\nu$ case since the Majorana nature of the neutrinos prevents absorbing two
of the phases).  In the following we will discuss how to measure many of the
parameters.  All are measurable in principle.  Once they are measured they
can be used to test any theory.  In practice, as always historically, some
measurements will be needed to formulate the underlying theory (e.g. to learn
how supersymmetry is broken and to compactify) and others will then test
approaches to doing that. 

Only 32 of these parameters are masses of mass eigenstates!  There are
four neutralinos, two charginos, four Higgs sector masses, three LH
sneutrinos, six each of charged sleptons, up squarks, and down
squarks, and the gluino.  We will examine the connections between the
soft masses and the mass eigenstates below.  Of the 32 masses, only
the gluino occurs directly in $L_{soft}$ --- the rest are all related
in complicated ways to $L_{soft}!$ One could add the gravitino with
its complex mass to the list of parameters.  Even the gluino mass gets
significant corrections that depend on squark masses.

Some of the ways these parameters contribute is to determining the
breaking of the EW symmetry and therefore to the Higgs potential, and
the masses and cross sections and decays of Higgs bosons, to the relic
density and annihilation and scattering of the LSP, to flavor changing
transitions because the rotations that diagonalize the fermion masses
will not in general diagonalize the squarks and sleptons, to
baryogenesis (which cannot be explained with only the CKM phase), to
superpartner masses and signatures at colliders, rare decays with
superpartner loops (e.g. $b\rightarrow s+\gamma),$ electric dipole and
magnetic dipole moments, and more.

\section{Connecting high and low scales}

Two of the most important successes of supersymmetry depend on
connecting the unification and EW scales.  We will not study this
topic in detail here since Martin covers it thoroughly, but we will
look at the aspects we need, particularly for the Higgs sector.  The
connection is through the logarithmic renormalization and running of
masses and couplings, with RGEs. In general we imagine the underlying
theory to be formulated at a high energy scale, while we need to
connect with experiment at the EW scale.  We can imagine running the
theory down (top-down) or running an effective Lagrangian determined
by data up (bottom-up).  It is necessary to calculate for all the
parameters of the superpotential and of the soft-breaking Lagrangian.
The RGEs are known for gauge couplings and for the superpotential
couplings to three loops, and to two loops for other parameters, for
the MSSM and its RH$\nu$ extension.  We will only look at one-loop
results since we are mainly focusing on pedagogical features.  An
interesting issue is that calculations must be done with
regularization and renormalization procedures that do not break
supersymmetry, and that is not straightforward.  How to do that is not
a solved problem in general, but it is understood through two loops
and more loops in particular cases, so in practice there is no
problem.

Since our ability to formulate a deeper theory will depend on deducing
from data the form of the theory at the unification scale, learning how
to convert EW data first into an effective theory at the weak scale, and
then into an effective theory at the unification scale, is in a sense
the major challenge for particle physics in the coming years.  There are
of course ambiguities in running to the higher scales.  Understanding
the uniqueness of the resulting high scale theory, and how to resolve
ambiguities as well as possible, is very important.

For the Higgs sector we need to examine the running of several of the
soft masses, whose RGEs follow.  The quantity $t$ is $\ln(Q/Q_{0}),$
where $Q$ is the scale and $Q_{0}$ a reference scale.

\begin{eqnarray}
16\pi^{2}dM_{H_{u}}^{2}/dt\approx 3X_{t}-6g_{2}^{2}\left|  M_{2}\right|  ^{2}
-\frac{6}{5}g_{1}^{2}\left|  M_{1}\right|  ^{2}
\end{eqnarray}

\begin{eqnarray}
16\pi^{2}dM_{H_{d}}^{2}/d\approx 3X_{b}+X_{\tau}-6g_{2}^{2}\left|  M_{2}\right|
^{2}-\frac{6}{5}\left|  M_{1}\right|  ^{2}
\end{eqnarray}

\noindent where

\begin{eqnarray}
X_{t,b}\approx 2\left|  Y_{t,b}\right|  ^{2}(M_{H_{u,d}}^{2}+m_{Q_{3}}^{2}+m_{\bar
{u}_{3},\bar{d}_{3}}^{2})+2\left|  a_{t,b}\right|  ^{2}
\end{eqnarray}

\noindent Note that $X_{t,b}$ are positive so $M_{H_{u,d}}^{2}$
decrease as they evolve toward the EW scale from a high scale, and
unless $\tan\beta$ is very large, $X_{t}$ is larger than $X_{b}.$ We
also need to look at just the leading behavior of the squark running,

\begin{eqnarray}
16\pi^{2}dM_{Q_{3}}^{2}/dt=X_{t}+X_{b}+...
\end{eqnarray}

\begin{eqnarray}
16\pi^{2}dM_{\bar{u}_{3}}^{2}/dt=2X_{t}+...
\end{eqnarray}

\begin{eqnarray}
16\pi^{2}dM_{\bar{d}_{3}}^{2}/dt=2X_{b}+...
\end{eqnarray}

\noindent Think back to the SM, where the coefficient (usually called
$\mu^{2}$ there but remember that $\mu$ is not the same as our $\mu)$
of $\phi^{2}$ in the Higgs potential must be negative to lead to
spontaneous symmetry breaking with the minimum of the potential away
from the origin.  Here $M_{H_{u}}^{2}$ plays the role, effectively, of
the SM $\mu^{2}.$ We see that because of the large $X_{t}$ the right
hand side of the equation for $M_{H_{u}}^{2}$ is indeed the largest,
and not only does $M_{H_{u}}^{2}$ decrease as it runs but the other
quantities run slower so they do not get vevs at the same time.  Thus
the theory naturally can lead to a derivation of the Higgs mechanism!
This is extremely important.  The theory could easily have had a form
where no Higgs vev formed, or where a Higgs vev could only form if
some squark also got a vev, which would violate charge and color
conservation.  The precise conditions for REWSB are somewhat more
subtle in supersymmetry --- $M_{H_{u} }^{2}$ does not actually need to
be negative, just smaller than $M_{H_{d}} ^{2},$ as we will see next.

\section{Radiative electroweak symmetry breaking (REWSB)}

The Higgs sector is the natural domain of supersymmetry.  The Higgs
mechanism \cite{11} occurs as the scale decreases from the more symmetric
high scale, with vacuum expectation values becoming non-zero somewhat
above the EW scale.  As we will see, the Higgs mechanism is
intricately tied up with supersymmetry and with supersymmetry breaking
--- there is no Higgs mechanism unless supersymmetry is broken.  This
should be contrasted with the other big issue of flavor physics, the
origin of the number of families and the differences between the
flavor and mass eigenstates, which is already in the structure of the
theory at the unification scale, as discussed above.  Supersymmetry
accommodates the flavor issues, and allows data to constrain them, but
supersymmetry can explain the Higgs physics with string boundary
conditions (we'll be more precise about that later).

Once we have the superpotential and $L_{soft}$ we can calculate the
scalar potential that determines the Higgs physics --- that is very
different from the SM case where one adds the scalar potential in by
hand.  The result is for the electrically neutral fields,

\begin{eqnarray}
V=\left|  \mu_{eff}\right|  ^{2}(\left|  H_{u}\right|  ^{2}+\left|
H_{d}\right|  ^{2})\hspace{1.6in}F
\end{eqnarray}

\begin{eqnarray}
+\frac{1}{8}(g_{1}^{2}+g_{2}^{2})(\left|  H_{u}\right|  ^{2}-\left|
H_{d}\right|  ^{2})\hspace{1.5in}D
\end{eqnarray}

\begin{eqnarray}
+m_{H_{u}}^{2}\left|  H_{u}\right|  ^{2}+m_{H_{d}}^{2}\left|  H_{d}\right|
^{2}-(bH_{u}H_{d}+c.c.).\hspace{.6in}soft
\end{eqnarray}

\noindent From now on again we will just write $\mu$ for $\mu_{eff}.$
Now we want to minimize this.  If it has a minimum away from the
origin vevs will be generated.  If we had included the charged scalars
we could use gauge invariance to rotate away any vev for (say)
$H_{u}^{+}.$ Then we would find that the minimization condition
$\partial V/\partial H_{d}^{-}=0$ implied that $\left\langle
H_{d}^{-}\right\rangle =0,$ so at the minimum electromagnetism is an
unbroken symmetry.  The only complex term in $V$ is $b.$ We can
redefine the phases of $H_{u},H_{d}$ to absorb the b phase, so we can
take $b$ as real and positive.  Then by inspection we will have a
minimum when the term with $b$ subtracts the most it can, so
$\left\langle H_{u}\right\rangle \left\langle H_{d}\right\rangle $
will be real and positive.  Since $H_{u},H_{d}$ have hypercharge
$\pm\frac{1}{2},$ we can use a hypercharge gauge transformation to
take the two vevs separately real and positive.  Therefore at the
tree level CP is conserved in the Higgs sector and we can choose the
mass eigenstates to have definite CP.

Writing $\partial V/\partial H_{u}=\partial V/\partial H_{d}=0$ one
finds that the condition for a minimum away from the origin is

\begin{eqnarray}
b^{2}>(\left|  \mu\right|  ^{2}+M_{H_{u}}^{2})(\left|  \mu\right|
^{2}+M_{H_{d}}^{2}).
\end{eqnarray}

\noindent So $M_{H_{u}}^{2}<0$ helps to generate EWSB but is not
necessary.  There is no EWSB if $b$ is too small, or if $\left|
\mu\right| ^{2}$ is too large.  For a valid theory we must also have the
potential bounded from below, which was automatic for the unbroken
theory but is not when the soft terms are included.  The quartic piece
in $V$ guarantees $V$ is bounded from below except along the so-called
D-flat direction $\left\langle H_{u}\right\rangle =\left\langle H_{d}
\right\rangle ,$ so we need the quadratic terms positive along that
direction, which implies

\begin{eqnarray}
2b<2\left|  \mu\right|  ^{2}+M_{H_{u}}^{2}+M_{H_{d}}^{2}.
\end{eqnarray}

\noindent Remarkably, the two conditions cannot be satisfied if
$M_{H_{u}}^{2}=M_{H_{d}}^{2}$, so the fact that $M_{H_{u}}^{2}$
runs more rapidly than $M_{H_{d}}^{2}$ is essential.  They also cannot
be satisfied if $M_{H_{u}}^2 = M_{H_{d}}^2 = 0$, i.e. if supersymmetry
is unbroken! 

We write $\left\langle H_{u,d}\right\rangle =$v$_{u,d}.$  Requiring
the Z mass be correct gives

\begin{eqnarray}
\rm{v}_{u}^{2}+\rm{v}_{d}^{2}=\rm{v}^{2}=\frac{2M_{Z}^{2}}{g_{1}
^{2}+g_{2}^{2}} \approx (174{\rm GeV})^{2}
\end{eqnarray}

\noindent and it is convenient to write

\begin{eqnarray}
\tan\beta=\rm{v}_{u}/\rm{v}_{d}.
\end{eqnarray}

\noindent Then v$_{u}=$v$\sin\beta,$ v$_{d}=$v$\cos\beta,$ and with
our conventions $0<\beta<\pi/2.$

With these definitions the minimization conditions can be written

\begin{eqnarray}
\left|  \mu\right|  ^{2}+M_{H_{d}}^{2}=b\tan\beta-\frac{1}{2}M_{Z}^{2}
\cos2\beta
\end{eqnarray}

\begin{eqnarray*}
\left|  \mu\right|  ^{2}+M_{H_{u}}^{2}=b\cot\beta+\frac{1}{2}M_{Z}^{2}
\cos2\beta.
\end{eqnarray*}

\noindent These satisfy the EWSB conditions.  They can be used (say) to
eliminate $b$ and $\left| \mu\right| ^{2}$ in terms of $\tan\beta$ and
$M_{Z}^{2}.$ Note the phase of $\mu$ is not determined.  These two
equations have a special status because they are the only two
equations of the entire theory that relate a measured quantity
($M_{Z}^{2})$ to soft parameters.  If the soft parameters are too
large, these equations would require very precise cancellations to
keep the Z mass correct.

We have two Higgs fields, each an SU(2) doublet of complex fields, so
8 real scalars.  Three of them are Nambu-Goldstone bosons that are
eaten by $W^{\pm},Z$ to become the longitudinal states of the vector
bosons, just as in the SM, so 5 remain as physical particles.  They
are usually classified as 3 neutral ones, $h,H,A,$ and a charged pair,
$H^{\pm}.$ The mass matrix is calculated from $V$ with
$M_{ij}^{2}=\frac{1}{2}\partial^{2}V/\partial\phi_{i}\partial\phi
_{j}$ where $\phi_{i,j}$ run over the 8 real scalars.  Then the
eigenvalue equation $\det\left| \lambda-M_{ij}^{2}\right| =0$
determines the mass eigenstates.  This splits into block diagonal
2$\times2 $ factors.  The factors for the charged states and the
neutral one in the basis
$({Im}H_{u},\;{Im}H_{d})$ each have one zero
eigenvalue, the Nambu-Goldstone bosons.  The two CP even neutrals can
mix, with mixing matrix

\begin{eqnarray}
\left(
\begin{array}
[c]{c}
h\\
H
\end{array}
\right)  =\sqrt{2}\left(
\begin{array}
[c]{cc}
\cos\alpha & -\sin\alpha\\
\sin\alpha & \cos\alpha
\end{array}
\right)  \left(
\begin{array}
[c]{c}
{Re}H_{u}-\rm{v}_{u}\\
{Re}H_{d}-\rm{v}_{d}
\end{array}
\right) .
\end{eqnarray}

\noindent The resulting tree level masses are

\begin{eqnarray}
m_{h,H}^{2}=\frac{m_{A}^{2}+M_{Z}^{2}}{2}\mp\frac{1}{2}\sqrt{(m_{A}^{2}
+M_{Z}^{2})^{2}-4m_{A}^{2}M_{Z}^{2}\cos^{2}2\beta},
\end{eqnarray}

\begin{eqnarray}
m_{A}^{2}=2b/\sin2\beta,
\end{eqnarray}

\begin{eqnarray}
m_{H^{\pm}}^{2}=m_{A}^{2}+M_{W^{\pm}}^{2}.
\end{eqnarray}

\noindent From eq. 54, one can see that if $m_{A}^{2}\rightarrow0$ then
$m_{h} ^{2}\rightarrow0,$ and if $m_{A}^{2}$ gets large then
$m_{h}^{2}\rightarrow0,$ so $m_{h}^{2}$ has a maximum.  A little
algebra shows the maximum is

\begin{eqnarray}
m_{h}^{tree}\leq\left|  \cos2\beta\right|  M_{Z},
\end{eqnarray}

\noindent where we have emphasized that this maximum does not include
radiative corrections.  This important result leads to the strongest
quantitative test of the existence of supersymmetry, that there must
exist a light Higgs boson.  If the gauge theory is extended to larger
gauge groups there are additional contributions to the tree level mass,
but they are bounded too.

There are also significant radiative corrections \cite{12}.  The Higgs
potential has contributions to the $h^{4}$ term from loops involving
top quarks and top squarks.  These are not small because the top
Yukawa coupling is of order unity and the top-Higgs coupling is
proportional to the top mass.  To include the effect one has to
calculate the contribution to the Higgs potential, reminimize, and
recalculate the mass matrix eigenvalues.  The result is

\begin{eqnarray}
m_{h}^{2}\lesssim \cos^{2}2\beta M_{Z}^{2}+\frac{3\alpha_{2}}{2\pi}\frac
{m_{t}^{4}}{m_{W}^{2}}\ln\frac{\tilde{m}_{t}^{2}}{M_{Z}^{2}} \approx
M_{Z}^{2}(1+\frac{1}{4}\ln\frac{\tilde{m}^{2}_t}{M_{Z}^{2}})
\end{eqnarray}

\noindent where the last equality uses $\left| \cos2\beta\right| =1,$
which is true for $\tan\beta\gtrsim 4.$ The contributions from two
loops have mainly been calculated and are small but not negligible.
This result shows that if $m_h \approx 115$ GeV, it is necessary that
the tree level term give essentially the full $M_Z$ contribution, so
$\cos^2 2\beta \approx 1$. 

If $\tan\beta$ is large the REWSB situation is more complicated.  Then
the top and bottom Yukawa couplings are approximately equal, so from the
RGEs [equations 39-41] we see that $M_{H_{u}}^{2},M_{H_{d} }^{2}$ run
together, and both can go negative, or the conditions [equations 48,49]
may not be satisfied.  The EWSB conditions can be rewritten [using
equation 55] so one condition is that

\begin{eqnarray}
2m_{A}^{2}\approx M_{H_{d}}^{2}-M_{H_{u}}^{2}-M_{Z}^{2}.
\end{eqnarray}

Experimentally, $m_{A}^{2} \gtrsim M_{Z}^{2}$ (or $A$ would have been
observed at LEP or the Tevatron), so the EWSB condition is that
$M_{H_{u}}^{2}$ must be smaller than $M_{H_{d}}^{2}$ by an amount
somewhat larger than $M_{Z}^{2}.$ That allows a narrow window, and
preferably the theory would not have to be finely adjusted to allow
the REWSB to occur.  Also, in this situation the other condition can
be written

\begin{eqnarray}
b\approx\frac{M_{H_{d}}^{2}-M_{H_{u}}^{2}}{\tan\beta}\sim\frac{M_{Z}^{2}}
{\tan\beta}\ll M_{Z}^2
\end{eqnarray}

\noindent when $\tan\beta$ is large, and this is a clear fine tuning \cite{31} 
since the natural scale for $b$ is of order the typical soft term,
presumably of order or somewhat larger than $M_{Z}^{2}.$ So REWSB is
possible with large $\tan \beta$ but it is necessary to explain why
this apparent fine tuning occurs.  The actual effects of increasing
$\tan\beta$ are complicated.  The b and $\tau$ Yukawas get larger, so
the top and stop and $m_{H_{u,d}}^{2}$ RGEs change.  $m_{H_{u,d}}^{2}$
get driven more negative, but the larger Yukawas decrease the stop
masses, which makes $m_{H_{u}}^{2}$ less negative, etc.

If $\tan\beta$ is large, theories with $M_{H_{u}}^{2}$ and $M_{H_{d}}^{2}$
split are then favored.  That could occur in the unification scale
formulation of the theory.  One possible way to get a splitting even if
$M_{H_{u}}^{2}, M_{H_{d}}^{2}$ start degenerate is via D-terms from
extending the gauge theory\cite{13}. D-terms arise whenever a U(1)
symmetry is broken.  Under certain circumstances their magnitude may be
of order the weak scale even though the U(1) symmetry is broken at a
high scale, and they can contribute if the superpartners are charged
under that U(1) symmetry.  If one looks at SO(10) breaking to
SU(5)$\times$U(1) and the breaking of this U(1), the soft masses are

\begin{eqnarray*}
%\begin{align*}
m_{Q}^{2} = m_{\bar{e}}^{2}=m_{\bar{u}}^{2}=m_{10}^{2}+m_{D}^{2}\\
m_{L}^{2} = m_{\bar{d}}^{2}=m_{5}^{2}-3m_{D}^{2}\\
m_{H_{d,u}}^{2} = m_{10}^{2}\pm2m_{D}^{2}.
\end{eqnarray*}

\noindent The main point for us is that $m_{H_{u}}^{2}$ and
$m_{H_{d}}^{2}$ are split.  The splitting affects the other masses, so
in principle \hspace{.1in}$m_{D}^2$ is accessible experimentally if
sufficiently many scalar masses can be measured.

Note that because $b$ is in $L_{soft}$ it is not protected by a
non-renormalization theorem.  So to have $b$ small at the weak scale
does not mean it is small at the unification scale.  It's RGE is

\begin{eqnarray*}
16\pi^{2}db/dt=b(3Y_{t}^{2}-3g_{2}^{2}+...)+\mu(6a_{t}Y_{t}+6g_{2}^{2}
M_{2}+...)
\end{eqnarray*}

\noindent so if it starts out at zero it is regenerated from the
second term, or alternatively cancellations can make it small at the
weak scale.  Such cancellations would look accidental or fine tuned if
one did not know the high scale theory, but the appropriate way to
view them would be as a clue to the high scale theory.  Similarly,
large $\tan\beta$ would presumably mean that one vev is approximately
zero at tree level and a small value is generated for it by radiative
corrections.  No theory is currently known that does that, but if an
appropriate symmetry can be found that does it will be a clue to the
high scale theory.

Before we leave Higgs physics we will derive one Feynman rule to
illustrate how that works. From above we write

\begin{eqnarray}
H_{d}=\rm{v}\cos\beta+\frac{1}{\sqrt{2}}(-h\sin\alpha+H\cos\alpha
+iA\sin\beta)
\end{eqnarray}

\begin{eqnarray*}
H_{u}=\rm{v}\sin\beta+\frac{1}{\sqrt{2}}(h\cos\alpha+H\sin\alpha+iA\cos
\beta).
\end{eqnarray*}

\noindent Then from the covariant derivative term there is the
Lagrangian contribution

\begin{eqnarray}
\frac{g_{2}^{2}}{\cos^{2}\theta_{W}}(\left|  H_{u}\right|  ^{2}+\left|
H_{d}\right|  ^{2})Z^{\mu}Z_{\mu}
\end{eqnarray}

\noindent so substituting this gives the $hZZ$ vertex

\begin{eqnarray}
\frac{g_{2}^{2}\rm{v}}{2\cos^{2}\theta_{W}}Z^{\mu}Z_{\mu}h(\sin\beta
\cos\alpha-\cos\beta\sin\alpha)=\frac{g_{2}M_{Z}}{\cos\theta_{W}}\sin
(\beta-\alpha)Z^{\mu}Z_{\mu}h.
\end{eqnarray}

\noindent Similar manipulations give the couplings

\begin{eqnarray}
\begin{tabular}
[c]{cccc}
\  & $h$ & $H$ & $A$\\
$\bar{t}t,$ $\bar{c}c,$ $\bar{u}u$ & $\cos\alpha/\sin\beta$ & $\sin\alpha
/\cos\beta$ & $\cot\beta$\\
$\bar{b}b,$ $\bar{\tau}\tau...$ & -$\sin\alpha/\cos\beta$ & $\cos\alpha
/\sin\beta$ & $\tan\beta$\\
$WW,ZZ$ & $\sin(\beta-\alpha)$ & $\cos(\beta-\alpha)$ & 0\\
$ZA$ & $\cos\left(  \beta-\alpha\right)  $ & $\sin\left(  \beta-\alpha\right)
$ & 0
\end{tabular}
\end{eqnarray}

\noindent The $ZAh$ and $ZHA$ vertices are non-zero, while the $Zhh$ and $ZHH$
vertices vanish; there is no tree level $ZW^{\pm}H^{\mp}$ vertex.

Finally, we note that in the supersymmetric limit where the soft
parameters become zero one has

\begin{eqnarray}
V=\left|  \mu\right|  ^{2}(\left|  H_{u}\right|  ^{2}+\left|  H_{d}\right|
^{2})+\frac{g_{1}^{2}+g_{2}^{2}}{2}(\left|  H_{u}\right|  ^{2}-\left|
H_{d}\right|  ^{2})
\end{eqnarray}

so the minimum is at $\mu=0,$ $H_{u}=H_{d};$ the latter implies
$\tan\beta$ =1.

\section{Yukawa couplings, $\tan\beta$ , and theoretical and
experimental constraints on $\tan\beta$}

It's important to understand how $\tan\beta$ originates, and what is
known about it.  At high scales the Higgs fields do not have vevs, so
$\tan\beta$ does not exist.  The superpotential contains information
about the quark and lepton masses through the Yukawa couplings.  As
the universe cools, at the EW phase transition vevs become non-zero
and one can define $\tan\beta=$v$_{u}/$v$_{d}$.  Then quark and lepton
masses become non-zero, $m_{q,l}=Y_{q,l}$v$_{u,d}.$

There are two values for $\tan\beta$ that are in a sense natural.  As
pointed out just above, the supersymmetric limit corresponds to
$\tan\beta$ =1.  Typically in string theories some Yukawa couplings
are of order gauge couplings, and others of order zero.  The large
couplings for each family are interpreted as the top, bottom, and tau
couplings.  If $Y_{t}\approx Y_{b}$ then $\tan\beta$ $\sim
m_{t}/m_{b}.$ Numerically this is of order 35, but a number of effects
could make it rather larger or smaller, e.g. the values of $m_{t}$ and
$m_{b}$ change considerably with scale, and with RGE running so
$m_{t}(M_{Z})/m_{b}(M_{Z})\sim50$.  Finally $\tan\beta$ is determined
at the minimum of the Higgs potential, and can be driven smaller.

There are theoretical limits on $\tan\beta$ arising from the
requirement that the theory stay perturbative at high scales
(remember, the evidence that the entire theory stays perturbative is
both the gauge coupling unification and the radiative EWSB).
Requiring that $Y_{t}=g_{2}m_{t} /\sqrt{2}M_{W}\sin\beta$ not diverge
puts a lower limit on $\sin\beta$ which corresponds to $\tan\beta$
$\gtrsim 1.2$ when done in the complete theory, and similarly
$Y_{b}=g_{2}m_{b}/\sqrt{2}M_{W}\cos\beta$ leads to $\tan\beta$
$\lesssim 60.$ This upper limit is probably reduced by REWSB.

There are no measurements of $\tan\beta$, and as I emphasize below it is
not possible to measure $\tan\beta$ at a hadron collider in
general. Perhaps we will be lucky and find ourselves in a part of
parameter space where such a measurement is possible, or more likely, a
combination of information from (say) $g_{\mu}-2$ and superpartner
masses will lead to at least useful constraints on $\tan\beta$.  LEP
experimental groups have claimed lower limits on $\tan\beta$ from the
absence of superpartner signals, but those are quite model dependent and
do not hold if phases are taken into account.  Similarly, there is a
real lower limit on $\tan\beta$ from the absence of a Higgs boson below
115 GeV, as explained above and in Section 22.  That limit is about 4 if
phases are not included, but lower when they are.

\section{In what sense does supersymmetry explain EWSB?}

Understanding the mechanism of EWSB, and its implications, is still
the central problem of particle physics.  Does supersymmetry indeed
explain it?  If so, the explanation depends on broken supersymmetry,
and we have seen that in the absence of supersymmetry breaking the EW
symmetry is not broken.  That's OK.  An explanation in terms of
supersymmetry moves us a step closer to the primary theory.
Historically we have learned to go a step at a time, steadily moving
toward more basic understanding.  If we think of supersymmetry as an
effective theory at the weak scale only, then we would expect the
sense in which it explains EWSB to be different from that we would
find if we think of low energy supersymmetry as the low energy
formulation of a high scale theory.  That is, top-motivated bottom-up
is different from bottom-up.  It should be emphasized that one could
have supersymmetry breaking without EWSB, but not EWSB without
supersymmetry breaking.

It may clarify the issues to first ask what needs explanation.  We can
explicitly list

(1) Why are there Higgs scalar fields, i.e. scalars that carry
SU(2)$\times$U(1) quantum numbers, at all?

(2) Why does the Higgs field get a non-zero vev?

(3) Why is the vev of order the EW scale instead of a high scale?

(4) Why does the Higgs interact differently with different particles,
in particularly different fermions?

Let us consider these questions.

At least scalars are naturally present in supersymmetric theories, and
generally carry EW quantum numbers, whereas in the SM scalars do not
otherwise occur.  If we connect to a high scale theory, some (most)
explicitly have SM-like Higgs fields, e.g. in the E$_{6}$
representation of heterotic string theories.  Basically as long as we
view supersymmetry as embedded in a high scale theory we will
typically have Higgs scalars present, though not in all possible
cases.  That in turn can point to the correct high scale theory.

We have seen that the RGE running naturally does explain the origin of
the Higgs vev if the soft-breaking terms and $\mu_{eff}$ are of order
the weak scale, and if one Yukawa coupling is of order the gauge
couplings. If we view the theory as a low energy effective theory we
have seen that we do not know why $\mu$ in the superpotential is zero,
but if we view the theory as embedded in a string theory then it is
natural to have $\mu=0$ in the superpotential.  We referred to this as
string boundary conditions.  Then how $\mu_{eff}$ is generated points
toward the correct high scale theory.  If $\mu_{eff}$ is of order the
weak scale then it is appropriate to explain the Higgs mechanism
{\it and} gauge coupling unification.  Similarly, the mechanism of
supersymmetry breaking has to give soft masses of order the weak scale
if supersymmetry explains (or, as some prefer to say, predicts)\textit{
}gauge coupling unification.

In a string theory, for example, we expect some Yukawa couplings to be
of order the gauge couplings.  We identify one of those with the top
quark.  Then the running of $M_{H_{u}}^{2}$ is fast and it is driven
negative, or decreases sufficiently, to imply the non-zero Higgs vev.
The relevant soft-breaking terms, particularly $M_{H_{u}}^{2}$ and
$M_{H_{d}}^{2} $ must be of order the weak scale.  The theory
accommodates different couplings for all the fermions. It does not
explain the numerical values of the masses, but allows them to be
different --- that is non-trivial.

So a complete explanation requires thinking of supersymmetry as embedded
in a deeper theory such as string theory $($so scalar fields exist in
the theory, and $\mu \approx 0,$ and the top Yukawa is of order 1$),$
and requires that the soft terms are of order the weak scale after
supersymmetry is broken.  If we only think of supersymmetry as a low
energy effective theory not all of these elements are present, so the
explanation is possible but incomplete.  It is not circular to impose
soft-breaking parameters of order the weak scale to explain the EWSB
since one is using supersymmetry breaking to explain EW breaking, which
is important progress --- that is how physics has increased
understanding for centuries.

It is also very important to note that the conditions on the existence
of Higgs and on $\mu$ and on the soft parameters are equally required
for the gauge coupling unification --- if they do not hold in a theory
then it will not exhibit gauge coupling unification.  The explanation
of EWSB requires in addition to the conditions for gauge coupling
unification only that there is a Yukawa coupling of order the gauge
couplings, i.e. a heavy top quark, which is a fact.

Perhaps it is amusing to note that two families are needed to have
both a heavy fermion so the EW symmetry is broken, and light fermions
that make up the actual world we are part of.  No reasons are yet
known why a third family is needed --- it is clear that CP violation
could have arisen from soft phases with two families, and does not
require the three family SM.

Now that we have developed some foundations we turn to applications in
several areas.

\section{Current and forthcoming Higgs physics}

There are two important pieces of information about Higgs physics that
both independently suggest it will not be too long before a confirmed
discovery.  But of course it is such an important question that solid
data is needed.

The first is the upper limit on $m_{h}$ from the global analysis of
precision LEP (or LEP + SLC +Tevatron) data \cite{14}.  Basically the
result is that there are a number of independent measurements of SM
observables, and every parameter needed to calculate at the observed
level of precision is measured except $m_{h}.$ So one can do a global
fit to the data and determine the range of values of $m_{h}$ for which
the fit is acceptable.  The result is that at 95\% C.L. $m_{h}$ should
be below about 200 GeV.  The precise value does not matter for us, and
because the data really determines $\ln m_{h}$ the sensitivity is
exponential so it moves around with small changes in input.  What is
important is that there is an upper limit.  The best fit is for a
central value of order 100 GeV, but the minimum is fairly broad.  The
analysis is done for a SM Higgs but is very similar for a supersymmetric
Higgs over most of the parameter space.

In physics an upper limit does not always imply there is something below
the upper limit.  Here the true limit is on a contribution to the
amplitude, and maybe it can be faked by other kinds of contributions
that mimic it.  But such contributions behave differently in other
settings, so they can be separated.  If one analyzes the possibilities
\cite{15} one finds that there is a real upper limit of order 450 GeV on
the Higgs mass, if (and only if) additional new physics is present in
the TeV region. That new physics or its effects could be detected at LHC
and/or a 500 GeV linear electron collider, and/or a higher intensity Z
factory (``giga-Z'') that accompanies a linear collider.  So the upper
limit gives us very powerful new information.  If no other new physics
(besides supersymmetry) occurs and conspires in just the required way
with the heavier Higgs state, the upper limit really is about 200 GeV.

The second new information is a possible signal from LEP \cite{16} in its
closing weeks for a Higgs boson with $m_{h}$=115 GeV.  The ALEPH
detector was the only group to do a blind analysis, and it is
technically a very strong detector, so its observation of about a
3$\sigma$ signal is important information.  It was not possible to run
LEP to get enough more data to confirm this signal.  Fortunately, its
properties are nearly optimal for early confirmation at the Tevatron,
since its mass is predicted, and cross section and branching ratio to
$b\bar{b}$ are large.  Less is required to confirm a signal in a
predicted mass bin than to find a signal of unknown mass, so less than
10 $fb^{-1}$ of integrated luminosity will be required if the LEP
signal is indeed correct.  If funding and the collider and the
detectors all work as planned confirming evidence for $h$ could come
in 2004.

Suppose the LEP $h$ is indeed real.  What have we learned \cite{17}?  Most
importantly, of course, that a point-like, fundamental Higgs boson
exists.  It is point-like because its production cross section is not
suppressed by structure effects.  It is a new kind of matter,
different from the century old matter particles and gauge bosons.  It
completes the SM, and points to how to extend the SM.  It confirms the
Higgs mechanism, since it is produced with the non-gauge-invariant
$ZZh$ vertex, which must originate in the gauge-invariant $ZZhh$
vertex with one $h$ having a vev.

The mass of $115$ GeV also tells us important information.  First, the
Higgs boson is not a purely SM one, since the potential energy would
be unbounded from below at that mass.  Basically the argument is that
one has to write the potential with quantum corrections, and the
corrections from fermion loops dominate because of the heavy top and
can be negative if $m_{h} $ is too small.  The SM potential is

\begin{eqnarray}
V(h)=-\mu^{2}h^{2}+\left\{  \lambda+\frac{3M_{Z}^{4}+6M_{W}^{4}+m_{h}
^{4}-12m_{t}^{4}}{64\pi^{2}\rm{v}^{4}}\ln(\textit{ \ })\right\}
h^{4},
\end{eqnarray}

\noindent where the argument of the $\ln$ is some function of the
masses larger than one.  In the usual way
$\lambda=m_{h}^{2}/2$v$^{2}.$ The second term in the brackets is
negative, so $\lambda$ and therefore $m_{h}$ has to be large enough.
The full argument has to include higher loops, thermal corrections, a
metastable universe rather than a totally stable one, etc., and
requires $m_{h}$ to be larger than about 125 GeV if $h$ can be a
purely SM Higgs boson.

Second, 115 GeV is an entirely reasonable value of $m_{h}$ for
supersymmetry, but only if $\tan\beta$ is constrained to be larger than
about 4.  That is because as described above, the tree level
contribution is proportional to $\left| \cos2\beta\right| $ and to get a
result as large as 115 it is necessary that $\left| \cos2\beta\right| $
be essentially unity, giving a lower limit on $\tan\beta$ of about 4 .
Even then the tree level can only contribute a maximum of $M_{Z}$ to
$m_{h}.$ The rest comes from the radiative corrections, mainly the top
loop.  Numerically one gets

\begin{eqnarray}
m_{h}^{2}\approx(91)^{2}+(40)^{2}\left\{\ln\frac{m_{\tilde{t}}^{2}}
{m_{t}^{2}}+...\right\}
\end{eqnarray}

\noindent where $m_{\tilde{t}}^{2}$ is an appropriate average of the
two stop mass eigenstates.  The second term must supply about 25 GeV,
which is quite reasonable.

The LEP signal, assuming it is correct, can only provide us a limited
amount of information since it only supplies two numbers, $m_{h}$ and
$\sigma\times BR.$ The full Higgs potential depends on at least 7
parameters \cite{18}, so none of them can be explicitly measured.  Because
the potential depends on the stop loops, it depends on the hermitean
stop mass matrix (equation 69 below).

Since the elements are complex, in general the loop contributions to
the Higgs potential will be complex, so the potential will have to be
re-minimized taking into account the possibility of a relative phase
between the Higgs vevs.  One can write

\begin{eqnarray}
H_{d}=\frac{1}{\sqrt{2}}\left(
\begin{array}
[c]{c}
\rm{v}_{d}+h_{d}+ia_{d}\\
h_{d}^{-}
\end{array}
\right)  ,\textit{ \ \ \ \ \ \ }H_{u}=\frac{e^{i\theta}}{\sqrt{2}}\left(
\begin{array}
[c]{c}
h_{u}^{+}\\
\rm{v}_{u}+h_{u}+ia_{u}
\end{array}
\right).
\end{eqnarray}

\noindent At the minimum of the potential it turns out that $\theta$
cannot be set to zero or absorbed by redefinitions.  The resulting
$\theta$ is a function of the phase of $\mu,$ $\phi_{\mu},$ and of the
phase(s) in $a_{t}$ (and of course of other parameters).  Thus $m_{h}$
and $\sigma_{h}\times BR(b\bar {b})$ are functions of the magnitudes
of $\mu$ and $a_{t},$ $m_{Q}^{2},$ $m_{u}^{2},$ $b,$ $\tan\beta$ , and
the physical phase(s) $\phi_{\mu} +\phi_{a_{t}}$ at least.  Since some
of these are matrices they can involve more than one parameter.  Also,
if $\tan\beta$ is large there will be important sbottom loops, and
chargino and neutralino loops can contribute.  So only in special
cases can data about the Higgs sector be inverted to measure
$\tan\beta$ and the soft parameters, and only then if there are at
least 7 observables.

If $\theta$ is significant then even and odd CP states mix and there
are 3 mixed neutral states which could all show up in the $b\bar{b}$
or $\gamma\gamma$ spectrum, and those spectra could show different
amounts of the three mass eigenstates.  Both cross section and BR for
the lightest state can be different from the SM and from the CP
conserving supersymmetry case.

One can check that the phase can be very important.  For example, if a
Higgs is observed at LEP and the Tevatron one can ask what region of
parameter space is consistent with a given mass and $\sigma_{h}\times
BR(b\bar{b}).$ The answer is significantly different, for example for
$\tan\beta$ , if the phase is included.  Or if no Higgs is observed
one can ask what region of parameters is excluded.  If the phase is
included the actual limit on $m_{h}$ is about 10\% lower than the
published limits from LEP, below 100 GeV.  Similarly, lower values of
$\tan\beta$ are allowed if phases are included than those reported by
LEP experimenters.

\section{The stop mass matrix}

Arranging the stop mass terms from the Lagrangian in the form

\begin{eqnarray*}
\left(  \tilde{t}_{L}^{\ast}\textit{ \ }\tilde{t}_{R}^{\ast}\right)
m_{\tilde{t}}^{2}\left(
\begin{array}
[c]{c}
\tilde{t}_{L}\\
\tilde{t}_{R}
\end{array}
\right)  ,
\end{eqnarray*}

\noindent the resulting Hermitean stop mass matrix is

\begin{eqnarray}
m_{\tilde{t}}^{2}=\left(
\begin{array}
[c]{cc}
m_{Q_{3}}^{2}+m_{t}^{2}+\Delta_{u} & \;\; {\rm v}(a_{t}\sin\beta-\mu Y_{t}
\cos\beta)\\
& m_{\bar{u}_{3}}^{2}+m_{t}^{2}+\Delta_{\bar{u}}
\end{array}
\right)  .
\end{eqnarray}

The $\Delta^{\prime}s$ are D-terms, from the $(\phi^{\ast}T\phi)^{2}$
piece of the Lagrangian ---
$\Delta_{u}=(\frac{1}{2}-\frac{2}{3}\sin^{2}\theta_{W} )\cos2\beta
M_{Z}^{2},$ $\Delta_{\bar{u}}=\frac{2}{3}\sin^{2}\theta_{W} \cos2\beta
M_{Z}^{2}.$ These EW D-terms are proportional to the T$_{3}$ and
hypercharge charges.  The pieces proportional to $\sin^{2}\theta_{W}$
come from the breaking of the U(1) symmetry and vanish if
$\sin^{2}\theta _{W}\rightarrow0.$ The $m_{t}^{2}$ comes from the
F-terms in the scalar potential,
$Y_{t}^{2}H_{u}^{0\ast}H_{u}^{0}\tilde{t}_{L}^{\ast}\tilde{t}_{L}$ and a
similar term for $\tilde{t}_{R},$ when the Higgs get vevs.  F-terms in
$V$ also give the term -$\mu
Y_{t}\tilde{t}^{\ast}\tilde{t}H_{d}^{0\ast}$ which gives the second term
in the 12 position when $H_{d}^{0}$ gets a vev.  The soft term
$a_{u}\tilde{t}^{\ast}\tilde{Q}_{3}H_{u}^{0}$ gives the first 12 term
when the Higgs gets a vev.  Similar mass matrices are written for all
the squarks and sleptons.  For the lighter ones the Yukawas and possibly
the trilinears are small, and the fermion masses are small, so only the
diagonal elements are probably large.  Each of the elements above is a
3$\times$3 matrix, so $m_{\tilde{t}}^{2}$ is a 6$\times$6 matrix.
$\rm{a}_t$ and $\mu$ and even $\rm{v}$ are in general complex.

\section{What can be measured in the Higgs sector?}

Assuming the LEP signal is indeed valid, as suggested particularly by
the ALEPH blind experiment, and it is confirmed at the Tevatron, what
can we eventually learn?  I will focus on the Tevatron and LHC since
they will be our only direct sources of Higgs information in the next
decade.  The Tevatron can use the $WWh,$ $ZZH$ channels.  In addition
once $m_{h}$ is known the inclusive channel, with about a $pb$ cross
section, can be used.  If the total cross section at the Tevatron for
Higgs production is 1.5 $pb$, and each detector gets 15 $fb^{-1}$ of
integrated luminosity, the total number of Higgs bosons produced is
about 45,000 in a known mass bin.  At some level it will be possible to
measure $g_{WWh}g_{b\bar{b}h}$ and g$_{ZZh}g_{b\bar{b}h}$ from $\sigma
\rm{xBR}$ for the WWh and ZZh channels, so their ratio tests whether $h$
couples to gauge bosons proportional to mass.  Once $m_{h}$ is known it
will be possible to see $h\rightarrow\tau\bar{\tau}$ in both inclusive
production and associated production with a $W$, and test if the
coupling to fermions is proportional to mass.  A similar test comes from
not seeing $h\rightarrow\mu\bar{\mu}$ $($or seeing a few events of this
mode since it should occur a bit below the 10$^{-3}$ level$).$ The
inclusive production is dominantly via a top loop so it measures
g$_{tth}$ indirectly, and this is complicated since superpartner loops
contribute as well as SM ones.  It may be possible to see the $t\bar
{t}h$ final state directly \cite{19}.  Since $BR(\gamma\gamma)$ is at
the 10$^{-3}$ level an observation or useful limit will be possible here
if the resolution is good enough.  All of these can give very important
tests of what the Higgs sector is telling us.

It is also interesting to ask if data can distinguish a SM Higgs from
a supersymmetric one, though most likely there will be signals of
superpartners as well as a Higgs signal so there will not be any
doubt. If $\tan\beta$ is large the ratio of $b\bar{b}$ to
$\tau\bar{\tau}$ is sensitive to supersymmetric-QCD effects and can
vary considerably from its tree level value \cite{20}.  The ratio of top to
bottom couplings is sensitive to ways in which the supersymmetric
Higgs sector varies from the SM one.  If $\tan\beta$ is large and
$m_{A}$ is less than about 150 GeV it is possible $A$ can be observed at
the Tevatron.  Altogether, the Tevatron may be a powerful Higgs
factory if it takes full advantage of its opportunities.  It is still
unlikely that there will be enough independent measurements at the
Tevatron to invert the equations relating the soft parameters and
$\tan\beta$ to observables.  The lighter stop mass eigenstate
$\tilde{t}_{1}$ may be observable at the Tevatron, and provide another
observable for the Higgs sector.

At LHC it is very hard to learn much about the lightest Higgs $h$ if
its mass is of order 115 GeV.  It will most likely be observed in the
inclusive production and decay to $\gamma\gamma,$ but observation in
the $\gamma\gamma$ mode does not tell us much about the Higgs physics
once the Higgs boson has been discovered, which will have occurred if
indeed $m_{h}\approx115$ GeV.  The $\gamma\gamma$ mode does not
demonstrate the Higgs mechanism is operating since it occurs for any
scalar boson. The SM does have a definite prediction for
BR($\gamma\gamma)$ from the top and $W$ loops, and superpartner loops
can be comparable, so a measurement would be very interesting. Note
that one cannot assume the $\gamma\gamma$ BR is known.

Maybe it will be possible to detect the $\tau\bar{\tau}$ mode at LHC
using WW fusion to produce h and tagging the quarks \cite{21}.  This mode
also confirms the non-gauge-invariant $WWh$ vertex.  Considerable
additional information about the Higgs sector may come from observing
the heavier Higgs masses and $\sigma\times BR,$ and the heavier stop.
Since $A\rightarrow\gamma\gamma$ but not to $ZZ,WW$ it may be possible
to see $A$ if it is not above the $t\bar{t}$ threshold. Decays of the
heavy Higgs to $\tau^{\prime}s$ are enhanced if $\tan\beta$ is
large. Note that one cannot assume only SM decays of h in analysis
since channels such as h$\rightarrow LSP+LSP$ are potentially open and
can have large BR since they are not suppressed by factors such as
$m_{b}^{2}/M_{W}^{2}.$ The combined data from the Tevatron and LHC may
provide enough observables to invert the Higgs sector, at least under
certain reasonable and checkable assumptions.

\section{Charginos}

The lightest superpartners are likely to be the
neutralinos and charginos, possibly the lighter stop, and the gluino.
Their mass matrices have entries from the higgsino-gaugino mixing once
the SU(2)$\times$U(1) symmetry is broken, so the mass eigenstates are
mixtures of the symmetry eigenstates.  When phases are neglected these
matrices are described in detail in many places so I will not repeat
that here.  However, it is worth looking at the most general case
including phases for several instructive reasons.  The chargino mass
matrix follows from the $L_{soft},$ in the wino-higgsino basis:

\begin{eqnarray}
M_{\widetilde{C}}=\left(
\begin{array}
[c]{cc}
M_{2}e^{i\phi_{2}} & \sqrt{2}M_{W}\sin\beta\\
\sqrt{2}M_{W}\cos\beta & \mu e^{i\phi_{\mu}}
\end{array}
\right)  .
\end{eqnarray}

The situation is actually more subtle --- this is a submatrix of the
actual chargino mass matrix, but this contains all the information ---
and the reader should see Martin or earlier reviews for details.
Also, the off-diagonal element can be complex too since it arises from
the last term in eq.14 when the Higgs gets a vev, and the vev can be
complex as explained above; I will just keep the phases of $M_{2}$ and
$\mu$ here. The masses of the mass eigenstates are the eigenvalues of
this matrix.  To diagonalize it one forms the hermitean matrix
$M^{\dag}M.$ The easiest way to see the main points are to write the
sums and products of the mass eigenstates,

\begin{eqnarray}
M_{\widetilde{C_{1}}}^{2}+M_{\widetilde{C}_{2}}^{2} = TrM_{\widetilde{C}
}^{\dag}M_{\widetilde{C}}=M_{2}^{2}+\mu^{2}+2M_{W}^{2},
\end{eqnarray}
\begin{eqnarray}
\hspace{-2.5in}M_{\widetilde{C}_{1}}^{2}M_{\widetilde{C}_{2}}^{2} = \det M_{\widetilde{C}
}^{\dag}M_{\widetilde{C}}
\end{eqnarray}
\begin{eqnarray*}
\hspace{.66in}= M_{2}^{2}\mu^{2}+2M_{W}^{4}\sin^{2}2\beta-2M_{W}
^{2}M_{2}\mu\sin2\beta\cos(\phi_{2}+\phi_{\mu})
\end{eqnarray*}

Experiments measure the masses of the mass eigenstates.  One thing to
note is that the masses depend on the phases $\phi_{2}$ and
$\phi_{\mu},$ even though there is no CP violation associated with the
masses.  Often it is implicitly assumed that phases can only be
measured by observing CP-violating effects, but we see that is not so.
The combination $\phi_{2}+\phi_{\mu}$ is a physical phase, invariant
under any reparameterization of phases, as much a basic parameter as
$\tan\beta$ or any soft mass.

If one wants to measure the soft masses, $\mu,$ $\tan\beta$ , $\phi
_{2}+\phi_{\mu}$ it is necessary to invert such equations.  Since
there are fewer observables than parameters to measure, additional
observables are needed.  One can measure the production cross sections
of the mass eigenstates.  But then additional parameters enter since
exchanges of sneutrinos (at an electron collider) or squarks (at a
hadron collider) contribute.  One can decide to neglect the additional
contributions, but then one is not really doing a measurement.  If one
``measures'' $\tan\beta$ from the above equations by setting the phase
to zero, as is usually done, the result is different from that which
would be obtained if the phase were not zero.  When the phases are
present the phenomenology, and any deduced results, can be quite
different.  We saw that for the Higgs sector above.  It is studied for
the chargino sector in ref. \cite{22}.  Similar arguments apply for the
neutralino mass matrix.

One implication of this analysis is that $\tan\beta$ is not in general
measurable at a hadron collider --- there are simply not enough
observables \cite{23}.  One can count them, and the equations never
converge.  Depending on what can be measured, by combining observables
from the chargino and neutralino sectors, and the Higgs sector, it may
be possible to invert the equations.  This is a very strong
argument\cite{23} for a lepton collider with a polarized beam, where
enough observables do exist if one is above the threshold for lighter
charginos and neutralinos, because measurements with different beam
polarizations (not possible at a hadron collider) double the number of
observables, and measurements with different beam energies (not possible
at a hadron collider) double them again.  The precise counting has to be
done carefully, and quadratic (and other) ambiguities and experimental
errors mean that one must do a thorough simulation\cite{23A} to be sure
of what is needed, but there appear to be sufficient observables to
measure the relevant parameters.  The issue of observing the fundamental
parameters of $L_{soft}$ is of course broader, as discussed in Section
17.  There are 33 masses in the MSSM including the gravitino, but 107
new parameters in $L_{soft}$ (including the gravitino).  The rest are
flavor rotation angles and phases.  Many can be measured by combining
data from a linear electron collider above the threshold for a few
superpartners and hadron colliders.  It is also necessary to include
flavor changing rare decays to measure the off-diagonal elements of the
sfermion mass matrices and the trilinear couplings.

\section{Neutralinos}

In a basis $\Psi^{0}=(\widetilde{B},\widetilde{W}_{3},\widetilde{H}
_{d},\widetilde{H}_{u})$ terms in the Lagrangian can be rearranged into
-$\frac{1}{2}(\Psi^{0})^{T}M_{\tilde{N}}\Psi^{0}$ with the symmetric

\begin{eqnarray*}
M_{\tilde{N}}=\left(
\begin{array}
[c]{cccc}
M_{1}e^{i\phi_{1}} & 0 & -\frac{g_{1}}{\sqrt{2}}H_{d}^{0\ast} & \frac{g_{1}
}{\sqrt{2}}H_{u}^{0\ast}\\
0 & M_{2}e^{i\phi_{2}} & \frac{g_{2}}{\sqrt{2}}H_{d}^{0\ast} & -\frac{g_{2}
}{\sqrt{2}}H_{u}^{0\ast}\\
&  & 0 & -\mu e^{i\phi_{\mu}}\\
&  &  & 0
\end{array}
\right)  .
\end{eqnarray*}

\noindent Although the elements are complex, this matrix can still be
diagonalized by a unitary transformation.  Its form in a basis
$\Psi^{\prime}=(\tilde{\gamma
},\widetilde{Z},\tilde{h}_{s},\tilde{h}_{a})$ is sometimes useful:

\[
M_{\tilde{N}}=\left(
\begin{array}
[c]{cccc}
M_{1}s_{W}^{2}+M_{2}c_{W}^{2} & \hspace{.2in}(M_{1}-M_{2})s_{W}c_{W} & 0 & 0\\
& M_{1}s_{W}^{2}+M_{2}c_{W}^{2} & M_{Z} & 0\\
&  & \mu\sin2\beta & -\mu\cos2\beta\\
&  &  & -\mu\sin2\beta
\end{array}
\right)  .
\]

\noindent If $M_{1}\approx M_{2}$ and/or if $\tan\beta$ is large (so
$\sin2\beta \approx0)$ this takes a simple form.

The lightest neutralino is the lightest eigenvalue of this, and may be
the LSP.  Its properties then determine the relic density of cold dark
matter (if the LSP is indeed the lightest neutralino).  It also largely
determines the collider signatures for supersymmetry.  It will be a
linear combination of the basis states,

\vspace{-.2in}
\begin{eqnarray*}
\widetilde{N}_{1}=\alpha\widetilde{B}+\beta\widetilde{W}_{3}+\gamma
\widetilde{H}_{d}+\delta\widetilde{H}_{u}
\end{eqnarray*}

\noindent with $\left|  \alpha\right|  ^{2}+\left|  \beta\right|  ^{2}+\left|
\gamma\right|  ^{2}+\left|  \delta\right|  ^{2}=1.$  

An interesting limit that is at least pedagogically instructive arises
if we take $M_{1}\approx M_{2}$ (at the EW scale) and $\tan\beta$
$\approx1,$ and $\mu<M_{Z}.$  Then
$\widetilde{N}_{1}\approx\tilde{h},$ where $\tilde
{h}=\tilde{h}_{d}\sin\beta+\tilde{h}_{u}\cos\beta,$ and
$M_{\tilde{N}_{1} }\approx\mu.$ 
$\widetilde{N}_{2}\approx\tilde{\gamma},$ with $M_{\widetilde
{N}_{2}}\approx M_{2},$ $\alpha\approx-\beta\approx-45^{\circ} $ so
$\cos(\alpha-\beta)$ and $\cos2\beta\rightarrow0.$  At tree level the
$Z\tilde{\gamma}\tilde{h},$ $h\tilde{\gamma}\tilde{h},$ and
$Z\tilde{h} \tilde{h}$ vertices vanish, and the dominant decay of the
second neutralino is
$\widetilde{N}_{2}\rightarrow\widetilde{N}_{1}+\gamma.$ 
$M_{\tilde{C}_{1} } \gtrsim M_{\widetilde{N}_{2}}.$

\section{Effects of phases}

The effects of phases have been considered much less than the masses.
As we saw above for charginos and the Higgs sector they affect not only
CP-violating observables but essentially all observables.  They can have
significant impacts in a variety of places, including $g_{\mu}-2$ ,
electric dipole moments (EDMs), CP violation in the K and B systems, the
baryon asymmetry of the universe, cold dark matter, superpartner
production cross sections and branching ratios, and rare decays.  We do
not have space to give a complete treatment, but only to make some
points about the importance of the observations and what they might
teach us about physics beyond the SM in general; while we focus to some
extent on the phases because they are usually not discussed, our concern
is relating them to the entire $L_{soft}.$

There are some experiments that suggest some of the phases are small,
mainly the neutron and electron EDMs.  On the other hand, we know that
the baryon asymmetry cannot be explained by the quark CKM phase, so some
other phase(s) are large, and the soft phases are good candidates.
Recently it has been argued that very large phases are needed if
baryogenesis occurs at the EW phase transition \cite{24}; see also
\cite{25}.  Further, there is no known symmetry or basic argument that
the soft phases in general should be small.  If the outcome of studying
how to measure them was to demonstrate that some were large that could
be very important because both compactification and supersymmetry
breaking would have to give such large phases.  The phase structure of
the effective soft Lagrangian at the weak scale and at the unification
scale are rather closely related, so it may be easier to deduce
information about the high scale phases from data than about high scale
parameters in general.  If the outcome of studying how to measure the
phases was to demonstrate that the phases were small that would tell us
different but very important results about the high scale theory.  It
would also greatly simplify analyzing weak scale physics, but that is
not sufficient reason to assume the phases are small.

\section{$g_{\mu}-2$}

In early 2001 it was reported that the anomalous magnetic moment of
the muon was larger than the SM prediction by a significant amount.
The experiment is now analyzing several times more data than the
original report was based on, and the SM theory is being reexamined.

Even if the effect disappears, it is worth considering $g - 2$
experiments, because in a supersymmetric world the entire anomalous
moment of any fermion vanishes if the supersymmetry is unbroken, so
magnetic moments are expected to be very sensitive to the presence of
low energy supersymmetry, and particularly of broken supersymmetry.  The
analysis can be done in a very general and model independent manner
\cite{26}, and illustrates nicely how one can say a great deal with
supersymmetry even though it seems to have a number of parameters.  So
it is also pedagogically interesting.  There are only two supersymmetric
contributions, a chargino-sneutrino loop and a smuon-neutralino
loop. One can see that starting from the complete theory, with no
assumptions beyond working in the MSSM, there are only 11 parameters
that can play a role out of the original set of over 100,

\vspace{-.2in}
\begin{eqnarray*}
\left|  M_{2}\right|  ,\left|  M_{1}\right|  ,\left|  \mu\right|  ,\left|
A_{\mu}\right|  ,m_{\tilde{\mu}_{L}},m_{\tilde{\mu}_{R}},m_{\tilde{\nu}}
,\tan\beta,\phi_{2}+\phi_{\mu},\phi_{1}+\phi_{\mu},\phi_{A}+\phi_{\mu
}.
\end{eqnarray*}

In the general case all 11 of them can be important, and the
experimental result will give a complicated constraint among them.
But if we ask about putting an upper limit on superpartner masses,
which would be of great interest, we can say more.  For larger masses
one can see that the chargino-sneutrino diagram dominates, and in
addition that it is proportional to $\tan\beta$; The $\tan\beta$
factor arises from the needed chirality flip on a chargino line.  Thus
only the magnitudes of $M_2$, and $\mu$, $\tan\beta$, $m_{\tilde{v}}$
and the phase enter in this limit.  If we illustrate the result by
assuming a common superpartner mass $\tilde{m}$ (just for pedagogical
reasons, not in the actual calculations), we find that

\begin{eqnarray}
a_{\mu}^{susy}/a_{\mu}^{SM}\approx\left(  \frac{100\textit{ GeV}}{\tilde{m}
}\right)  ^{2}\tan\beta\cos(\phi_{2}+\phi_{\mu}).
\end{eqnarray}

Further, to put upper limits on the masses we can take the phase to be
zero since it turns out to enter only in the above form under these
assumptions (for the general case see \cite{27}).  And if we express results
in terms of the lighter chargino mass rather than $M_{2}$ and $\mu$ we
can eliminate one parameter; for a given chargino mass there will be
ranges of $M_{2}$ and $\mu.$ So we are down to three parameters, with
no uncontrolled approximations or assumptions.  We will not focus on
details of the data here since the new data in 2002 will in any
case require a new analysis.  If the effect persists there will be
significant upper limits on the superpartner masses.  Note the
relevant physical phase here is $\phi_{2}+\phi_{\mu}.$ 

It is interesting to consider the supersymmetry limit so the
supersymmetric SM contribution vanishes.  In that limit the two
lighter neutralino masses vanish, and their contribution cancels the
photon contribution, the two heavier neutralino masses become $M_{Z}$
and their contribution cancels that of the $Z,$ and the two charginos
have $M_{W}$ and cancel the $W$ contribution.  Since the chargino has
a sign opposite to that of the $W$ in the supersymmetric limit but the
same sign for the broken supersymmetry physical situation it is
important to check that indeed the piece proportional to $\tan\beta$
does change sign as needed.

\section{Electric dipole moments}

In the SM electric dipole moments are unobservably small, of order
10$^{-33}e$ cm.  That is basically because they are intrinsically
CP-violating quantities, and for CP violation to occur in the SM it is
necessary for all three families to affect the quantity in question.
 Otherwise one could rotate the CKM matrix in such a way that the phase did
not occur in the elements that contributed.  So it must be at least a
two-loop suppression.  There must also be a factor of the electron or quark
mass because of a chirality flip, with the scale being of order $M_{W}.$  In
addition there is a GIM suppression.  Interpreting results will be
complicated because the neutron, and any nuclear EDM, can have a contribution
from strong CP violation, while the electron can only feel effects from EW
interactions.  

Naively, the EDM is the imaginary part of a magnetic moment operator,
and the real part is the magnetic moment.  So EDMs can arise from the
same diagrams as $g_{\mu}-2$ , but for the electron and for quarks (in
neutrons).  It is more complicated in reality because the part of the
amplitude that has an imaginary part may not give the dominant
contribution to the magnetic moment.  It has been known for a long
time that if the soft phases were of order unity and if all
contributions were independent, then the supersymmetry contributions
to EDMs are too large by a factor of order 50.  However, over a
significant part of parameter space various contributions can cancel.
Some of that cancellation is generic, e.g. between chargino and
neutralino in the electron EDM because of the relative minus sign in
eq.32.  The smallness of EDMs may be telling us that the soft phases
are small.  Then we need to find out why they are small.  Or it may be
telling us that cancellations do occur.  Cancellations look fine tuned
from the point of view of the low energy theory, but small phases look
fine tuned too.  Relations among soft parameters in the high scale
theory will look fine-tuned in the low scale theory if we do not know
the origin of those relations.  If $\tan\beta$ is very large
cancellations become unlikely since the chargino contribution will
dominate the eEDM just as it does for $g_{\mu}-2$ , but if $\tan\beta$
is of order 4-5 the situation has to be studied carefully.

\section{Measuring phases at hadron colliders}

Phases, as well as soft masses, can affect distributions at colliders.
 We briefly illustrate that here for an oversimplified model
\cite{28}.  Consider gluino production at a hadron collider.  The
Lagrangian contains a term

\begin{eqnarray}
M_{3}e^{i\phi_{3}}\lambda_{\tilde{g}}\lambda_{\tilde{g}}+c.c.
\end{eqnarray}

\noindent It is convenient to redefine the fields so the phase is
shifted from the masses to the vertex, so one can write
$\psi_{\tilde{g}}=e^{i\phi_{3} /2}\lambda_{\tilde{g}}.$ Then writing
the Lagrangian in terms of $\psi$ the vertices $q\tilde{q}\tilde{g}$
get factors $e^{\pm i\phi_{3}/2}.$ The production cross sections for
gluinos, for example from $q+\bar{q} \rightarrow\tilde{g}+\tilde{g}$
by squark exchange, have factors $e^{+i\phi_{3}/2}$ at one vertex and
$e^{-i\phi_{3}/2}$ at the other, so they do not depend on the phase.
That is clear from general principles since $\phi_{3} $ is not by
itself a physical, reparameterization-invariant phase.  But gluinos
always decay, and for example in the decay $\tilde{g}\rightarrow
q+\bar{q}+\tilde{B}$ mediated by squarks there is a factor of
$e^{i\phi_{3}/2} $ at the $q\tilde{q}\tilde{g}$ vertex and a factor
$e^{i\phi_{1}/2}$ at the $q\tilde{q}\tilde{B}$ vertex, so the rate
depends on the physical relative phase $\phi_{3}-\phi_{1}.$ In general
it is more complicated with all the relative phases of the neutralino
mass matrix entering.  In this simple example the experimental
distribution in Bino energy is

\begin{eqnarray}
d\sigma/dx\sim m_{\tilde{g}}^{4}\left(  \frac{1}{\tilde{m}_{L}^{4}}+\frac
{1}{\tilde{m}_{R}^{4}}\right)\times  
\end{eqnarray}
\begin{eqnarray*}
[x-4x^{2}/3-2y^{2}/3+y(1-2x+y^{2})\cos(\phi_{3}-\phi_{1})]
\end{eqnarray*}

\noindent where $x=E_{\tilde{B}}m_{\tilde{g}}$ and
$y=m_{\tilde{B}}m_{\tilde{g}}.$ Other distributions are also affected.
If one tries to obtain information from gluino decay distributions
without taking phases into account the answers will be misleading if
the phases are not small.  The same result is of course true for many
superpartner decays.  It is important to realize that the same phases
are appearing here as appear for example in studying $\varepsilon$ and
$\varepsilon^{\prime}$ in the kaon system or in $b\rightarrow
s+\gamma.$

\section{LSP cold dark matter}

If it is stable, the LSP is a good candidate for the cold dark matter
of the universe.  Historically, it is worth noting that this was
noticed before we knew that non-baryonic dark matter was needed to
understand large scale structure.  It is a prediction of
supersymmetry.  We discussed above why we expected R-parity or a
similar symmetry to hold, with the stability of the LSP as one of its
consequences.  Then the basic argument is simple.  As the universe
cools, soon after the EW phase transition all particles have decayed
except photons, $e^{\pm},u^{\pm},d^{\pm},$ neutrinos, and LSPs.  The
quarks form baryons, which join with electrons to make atoms.  The
relic density of all but LSPs is known to be $\Omega_{SM}<0.05,$ while
$\Omega_{matter}\approx1/3.$ The LSPs annihilate as the universe
cools, with a typical annihilation cross section
$\sigma_{ann}\sim\rho_{LSP}G_{F}^{2} E^{2},$ and in the early universe
$E\sim T.$ The expansion rate is governed by the Hubble parameter
$H\sim T^{2}/M_{Pl}.$ The LSPs freeze out and stop annihilating when
their mean free collision path is of order the horizon, so
$\sigma_{ann}\sim H.$ This gives a density
$\rho_{LSP}\sim1/M_{Pl}G_{F} ^{2}\sim10^{-9}$ GeV$^{3}.$ At freeze-out
$T \sim 1$ GeV, and $\rho _{\gamma}$ is of order the entropy $S\sim
T^{3}\sim1$ GeV$^{3},$ so $\rho_{LSP}/\rho_{\gamma}\sim10^{-9},$
similar to the density of baryons.  Thus
$\Omega_{LSP}\sim(M_{LSP}/M_{proton})\Omega_{baryon}.$ Quantitative
calculations in many models confirm this.

But the actual calculations of the relic density depend on several
soft parameters such as masses of sleptons and gauginos, and also on
$\tan\beta$ and on soft phases.  In the absence of measurements or a
theory that can convincingly determine all of these, we cannot in fact
say more than that qualitatively the LSP is a good candidate, even if
WIMPs are apparently discovered.  Since we have argued above that in
practice it is unlikely that $\tan\beta$ will be measured accurately
at hadron colliders (though we may be lucky with $g_{\mu}-2$ plus
hadron colliders), it may be difficult to compute $\Omega_{matter}$
accurately even after LSPs are detected.  It should be emphasized that
detection of LSPs is not sufficient to argue they are actually
providing the cold dark matter \cite{29} --- LSPs could be detected in
direct experiments scattering off nuclei, and in space based searches,
and at colliders even if $\Omega_{LSP}\lesssim 0.05$.  Alternatively,
they could make up the CDM even though they were not detected in
direct and space based experiments.  

Further, in recent years it has come to be understood that LSPs may be
produced dominantly by processes that are not in thermal equilibrium
rather than the equilibrium process described above.  In that case the
relic density is not so simply connected to the LSP nature.

\section{Comments on relating CP violation and string theory; could the
CKM phase be small?}

Where does CP violation originate?  Can the pattern of CP phenomena
give us important clues to formulating and testing string theory?
Very little work has been done about the fundamental origins of CP
violation.  In 1985 Strominger and Witten discussed how to define CP
transformations in string theory, and in 1993 Dine, Leigh, and
McIntyre argued that CP was a gauge symmetry in string theory, for
both strong and EW CP violation.  As a gauge symmetry it could not be
broken explicitly, perturbatively. or non-perturbatively.  More
recently Bailin et al, Dent, Geidt, and Lebedev have discussed aspects
of this question.  Little thought has been given to CP violation in
D-brane worlds, Type IIB theories with SM particles as Type I open
strings, and so on.

From the point of view of connecting to the observable world, however
the CP violation originates it will appear as phases in either the
Yukawa couplings in the superpotential, or as phases in $L_{soft}.$
Any theory for CP violation will produce characteristic patterns of
such phases.  So if we could measure those phases perhaps we would
have rather direct information about such questions as moduli
dependence of Yukawas, supersymmetry breaking and transmission, vevs
of moduli and the dilaton, and the compactification manifold.

If one begins with a string theory including proposed solutions to how
to compactify, and to break supersymmetry, the connection to the
observable world is first made by writing down the Kahler potential,
gauge kinetic function, and superpotential,
$W=Y_{\alpha\beta\gamma}\phi_{\alpha}\phi _{\beta}\phi_{\gamma}.$ Then
$L_{soft}$ is calculated for the assumed approach to supersymmetry
breaking, etc.  The trilinear terms, for example, are linear
combinations of the Yukawas and derivatives of the Yukawas with respect
to moduli fields.  So if the Yukawas have large phases it seems likely
the trilinear terms will also have large phases.  On the other hand,
phases could enter the trilinears through the Kahler potential even if
they were not present in the Yukawas.  Recalling that the quark CKM
phase is unable to provide the CP violation needed for the baryon
asymmetry, it is interesting to consider the possibility that all CP
violation originates in the soft phases.  It is possible to describe CP
violation in the kaon and B systems with only soft phases \cite{30}.

Phenomenologically there are a number of ways that soft phases could
be shown to be large.  One is observing an eEDM.  The nEDM is not so
simple to interpret since it could arise from strong CP violation, but
perhaps the relative size of the nEDM and HgEDM could show the effect
of soft phases.  The Higgs sector could show phase effects, as could
superpartner masses, production cross sections, and decay BR.  The
size of $K_{L}\rightarrow \pi^{0}\nu\bar{\nu}$ could deviate from the
SM prediction.  It is much harder to demonstrate that
$\delta_{CKM}\neq0.$

\section{Phases (and flavor structure) of $L_{soft}$}

The soft-breaking Lagrangian has, as we have seen, many phases, and
interesting and potentially important flavor structure.  Few top-down
models, e.g. string based models, have studied or even looked at the
phase and flavor structure.  There is and will be much more data on
these topics, and there should be much more theoretical analysis of
them.  We have looked a little at string motivated models that give
interesting phase structure.  There is some work on this by Bailin et
al for the heterotic string.  Following the framework of Ibanez,
Munoz, and Rigolin\cite{37}, we have looked at how the phases emerge in
some D-brane models \cite{32}.

If one embeds the MSSM on one brane, usually the gaugino masses $M_{i}$
all have the same phase, and using the freedom from a U(1) symmetry as
one can rotate that phase away.  An interesting structure emerges if one
embeds the SM gauge groups on two intersecting branes.  We studied the
simplest case with SU(2) on one brane, and SU(3)$\times$U(1) on the
other.  While we did not try to derive such a structure from an actual
compactification, it is known that explicit compactifications of
intersecting branes exist, and that open strings connecting D-branes
intersecting at non-vanishing angles lead to theories with chiral
fermions, so it is plausible that such a model can exist.  We follow
Ibanez et al in assuming the supersymmetry breaking occurs in a hidden
sector, and is transmitted by complex F-term moduli vevs to the
superpartners.  Then this model gives for soft terms

\begin{eqnarray}
M_{1}=M_{3}=-A_{t}\sim e^{i\alpha_{1}},\textit{ \ \ \ \ \ \ }M_{2}\sim
e^{-i\alpha_{2}},
\end{eqnarray}

\noindent and all the other soft terms are real.  One important lesson
is that such a theory has only 9 parameters --- the many parameters of
$L_{soft}$ have been reduced by the theory down to this number.  They
are

\begin{eqnarray}
\alpha_{2}-\alpha_{1},m_{3/2},\tan\beta,\left|  \mu\right|  ,\left|
A_{t}\right|  ,\phi_{\mu},X_{1},X_{2},X_{3}.
\end{eqnarray}

Here only the relative phase of the moduli vevs enters, $m_{3/2}$ is
the gravitino mass and sets the overall mass scale, and the $X_{i}$
are measures of the relative importance of different moduli.  The
$X_{i}$ could be measured, in which case they would tell us about the
structure of the theory, and/or they could be computed in a good
theory.  Measuring the string-based parameters here would teach us
about formulating and testing string theory.  Any theory will have
relations among the soft parameters so the actual number of parameters
is far smaller than the full number of $L_{soft}.$ This number could
be reduced further by some assumptions.  Also, not all of them will
contribute in any given process, as we have seen.  The resulting
theory can be used to simultaneously study collider physics and LSP
cold dark matter as is usual, and also CP violation.  An extended
version of the model \cite{32B} can also address flavor issues.

In this model one can illustrate how results of the low energy theory
can appear fine-tuned and somewhat arbitrary because they are not
apparently due to a symmetry when they originate in dynamics that
occur at the high scale and are hidden at the low energy scale.  If
the gluino-gluon box diagram indeed explains direct CP violation in
the kaon system, then one needs a certain phase relation to hold,

\begin{eqnarray}
\arg\left(  \phi_{A_{sd}}M_{3}^{\ast}\right)  \approx10^{-2},
\end{eqnarray}

\noindent which seems fine-tuned.  But as we saw in eq.76, $M_{3}$ and
the elements of $A$ have the same phase in this D-brane based theory,
and so the quantity in eq.78 is zero at the high scale.  Since the
phases of $M_{3}$ and of $A$ run differently, a small phase is
generated at the low scale.  While we are not arguing this is the
actual explanation for $\varepsilon_{K}^{\prime},$ it does nicely
illustrate how such phases could be related by an underlying theory
yet not follow from any low energy symmetry.

\section{Direct evidence for superpartners? --- at the tevatron?}

So far all the evidence for low energy supersymmetry is indirect.
Although the evidence is strong, it could in principle be a series of
coincidences.  More indirect evidence could come soon from improved
$g_{\mu }-2$ , other rare decays, b-factories, proton decay, CDM
detection.  But finally it will be necessary to directly observe
superpartners, and to show they are indeed superpartners.  That could
first happen at the Tevatron collider.

Indeed, as we discussed earlier, if supersymmetry is really the
explanation for EWSB then the soft masses should be of order $M_{Z},$
and the cross sections for their production are typical EW ones, or
larger for gluinos, so superpartners should be produced in significant
quantities at the Tevatron collider that has just begun to take data
after a six year upgrade in luminosity.  Assuming the luminosity and
the detectors are good enough to separate signals from backgrounds, if
direct evidence for superpartners does not emerge at the Tevatron then
either nature does not have low energy supersymmetry or there is
something completely missing from our understanding of low energy
supersymmetry.  There is no other hint of such a gap in our understanding.
Thus if superpartners do not appear at the Tevatron many people will
wait until LHC has taken data to be convinced nature is not
supersymmetric, but it is unlikely that superpartners could be found
at the LHC if they are not first found at the Tevatron.  So let us
examine how they are likely to appear at the Tevatron.

Accepting that supersymmetry explains EWSB, we expect the gluinos,
neutralinos, and charginos to be rather light.  The lighter stop may
be light as well.  Sleptons may also be light though there is somewhat
less motivation for that.  We can list a number of channels and look
at the signatures for each of them.  Almost all cases require a very
good understanding of the SM events that resemble the possible
signals, both in magnitude (given the detector efficiencies) and the
distributions.  Missing transverse energy will be denoted by $\noe_T$.
It is reasonable to expect the Tevatron to have an integrated
luminosity of 2 $fb^{-1}$ per detector by sometime in 2004, and 15
$fb^{-1}$ by sometime in 2007.  Until we know the ordering of the
superpartner masses we have to consider a number of alternative decays
of $\widetilde{N}_{2},$ $\widetilde{C}_{1},$ $\tilde{t} _{1},$
$\tilde{g},$ etc. \cite{33}.

$\bullet$ $\widetilde{N}_{1}+\widetilde{N}_{1}$ 

This channel is very hard to tag at a hadron collider.

$\bullet$ $\widetilde{N}_{1}+\widetilde{N}_{2,3}$

These channels can be produced through an s-channel $Z$ or a t-channel
squark exchange.  The signatures depend considerably on the character of
$\widetilde{N}_{2},$ $\widetilde{N}_{3}.$ $\widetilde{N}_{1}$ escapes.
If $\widetilde{N}_{2}$ has a large coupling to $\widetilde{N}_{1}+Z$
(for real or virtual $Z$) then the $\widetilde{N}_{1}$ will escape and
the $Z$ will decay to $e$ or $\mu$ pairs each 3$\%$ of the time, so the
event will have missing energy and a prompt lepton pair.  There will
also be tau pairs and jet pairs, but those are somewhat harder to
identify.  Or, perhaps $\widetilde{N}_{2}$ is mainly photino and
$\widetilde{N}_{1}$ mainly higgsino, in which case there is a large BR
for $\widetilde{N}_{2}\rightarrow\widetilde{N}_{1}+\gamma$ and the
signature of $\widetilde{N}_{2}$ is one prompt $\gamma$ and missing
energy.  The production cross section can depend significantly on the
wave functions of $\widetilde{N}_{1},\widetilde{N}_{2}.$ If the cross
section is small for $\widetilde{N}_{1}+\widetilde{N}_{2}$ it is likely
to be larger for $\widetilde{N}_{1}+\widetilde{N}_{3}.$ Most cross
sections for lighter channels will be larger than about 50 $fb$, which
corresponds to 200 events (not including BR) for an integrated
luminosity of 2 $fb^{-1}$ per detector.

$\bullet$ $\widetilde{N}_{1}+\widetilde{C}_{1}$

These states are produced through s-channel $W^{\pm}$ or t-channel
squarks.  The $\widetilde{N}_{1}$ escapes, so the signature comes from
the $\widetilde{C}_{1}$ decay, which depends on the relative sizes of
masses, but is most often $\widetilde{C}_{1}\rightarrow l^{\pm}+$$\noe_T$.
This is the signature if sleptons are lighter than charginos
($\widetilde{C} _{1}\rightarrow\tilde{l}^{\pm}+\nu,$ followed by
$\tilde{l}^{\pm}\rightarrow l^{\pm}+\widetilde{N}_{1}),$ or if
sneutrinos are lighter than charginos by a similar chain, or by a
three-body decay ($\widetilde{C}_{1}\rightarrow
\widetilde{N}_{1}+virtual$ $W,$ $W\rightarrow l^{\pm}+\nu).$ But it is
not guaranteed --- for example if stops are lighter than charginos the
dominant decay could be $\widetilde{C}_{1}\rightarrow\tilde{t}+b.$
 In the case where the lepton dominates the event signature is then
$l^{\pm}+$$\noe_T$ , so it is necessary to find an excess in this
channel.  Compared to the SM sources of such events the supersymmetry
ones will have no prompt hadronic jets, and different distributions
for the lepton energy and for the missing transverse energy.

$\bullet$ $\widetilde{N}_{2}+\widetilde{C}_{1}$

If $\widetilde{N}_{2}$ decays via a Z to $\widetilde{N}_{1}+l^{+}
+l^{-}$ and $\widetilde{C}_{1}$ decays to $\widetilde{N}_{1}+l^{\pm},$
this channel gives the well-known ``tri-lepton'' signature, three
charged leptons, $\noe_T$ , and no prompt jets, which may be relatively
easy to separate from SM background.  But it may be that
$\widetilde{N}_{2}\rightarrow\widetilde{N}_{1}+\gamma,$ so the signature
may be $l^{\pm}+\gamma+$$\noe_T$.

$\bullet$ $\tilde{l}^{+}+\tilde{l}^{-}$

Sleptons may be light enough to be produced in pairs.  Depending on
masses, they could decay via $\tilde{l}^{\pm}\rightarrow
l^{\pm}+\widetilde {N}_{1},$ $\widetilde{C}_{1}+\nu,$ $W+\tilde{\nu}.$
If $\widetilde{N}_{1}$ is mainly higgsino decays to it are suppressed
by lepton mass factors, so $\tilde{l}^{\pm}\rightarrow
l^{\pm}+\widetilde{N}_{2}$ may dominate, followed by
$\widetilde{N}_{2}\rightarrow\widetilde{N}_{1}+\gamma$ \cite{34}.

For a complete treatment one should list all the related channels, and
combine those that can lead to similar signatures.  The total sample may
be dominated by one channel but have significant contributions from
others, etc.  It should also be emphasized that the so-called
``backgrounds'' are not junk backgrounds that cannot be calculated, but
from SM events whose rates and distributions can be completely
understood.  Determining these background rates is essential to
identifying a signal and to identifying new physics, and requires
powerful tools in the form of simulation programs, which in turn require
considerable expertise to use correctly.  The total production cross
section for all neutralino and chargino channels at the Tevatron
collider is expected to be between 0.1 and 10 $pb$, depending on how
light the superpartners are, so even in the worst case there should be
several hundred events in the two detectors.  If the cross sections are
on the low side it will require combining inclusive signatures to
demonstrate new physics has been observed.

$\bullet$ gluinos can be produced via several channels,
$\tilde{g}+\tilde{g},$
$\tilde{g}+\widetilde{C}_{1},\tilde{g}+\widetilde{N_{1}},$ etc. 

If supersymmetry indeed explains EWSB it would be surprising if the
gluino were heavier than about 500 GeV, as argued above.  Then the
total cross section for its production should be large enough to
observe it at the Tevatron.  If all its decays are three-body,
e.g. $\tilde{g}\rightarrow \tilde{q}+\bar{q}$ followed by
$\tilde{q}\rightarrow q+\widetilde{C}_{1},$ etc, then the signature
has energetic jets, $\noe_T$, and sometimes charged leptons.  There are
two channels that are particularly interesting and not unlikely to
occur --- if $t+\tilde{t}$ or $b+\tilde{b}$ are lighter than
$\tilde{g}$ then they will dominate because they are two-body.  The
signatures can then be quite different, with mostly b and c jets, and
smaller multiplicity.

Gluinos and neutralinos are normally Majorana particles.  Therefore they
can decay either as particle or antiparticle.  If, for example, a decay
path $\tilde {g}\rightarrow\bar{t}(\rightarrow W^{-}\bar{b})+\tilde{t}$
occurs, with $W^{-}\rightarrow e^{-}\nu,$ there is an equal probability
for $\tilde {g}\rightarrow e^{+}+....$ Then a pair of gluinos can with
equal probability give same-sign or opposite sign dileptons!  The same
result holds for any way of tagging the electric charge --- we just
focus on leptons since their charge is easiest to identify.  The same
result holds for neutralinos.  The SM allows no way to get prompt
isolated same-sign leptons, so any observation of such events is a
signal beyond the SM, and very likely a strong indication of
supersymmetry.

$\bullet$ Stops can be rather light, so they should be looked for very
seriously.  They can be pair-produced via gluons, with a cross
section that is about 1/8 of the top pair cross section.  It is
smaller because of a p-wave threshold suppression for scalars, and a
factor of 4 suppression for the number of spin states.  They could
also be produced in top decays if they were lighter than
$m_{t}-M_{\tilde{N}_{1}},$ and in gluino decays if they are lighter
than $m_{\tilde{g}}-m_{t},$ which is not at all unlikely.  Their
obvious decay is $\tilde{t}\rightarrow\widetilde{C}+b,$ which will
indeed dominate if $m_{\tilde{t}}>m_{\widetilde{C}}.$  If this
relation does not hold, it may still dominate as a virtual decay,
followed by $\widetilde{C}$ real or virtual decay (say to
$W+\widetilde{N}_{1}),$ in which case the final state is 4-body after
$W$ decays, and suppressed by 4-body phase space.  That may allow the
one-loop decay $\tilde{t}\rightarrow c+\widetilde{N}_{1}$ to dominate
stop decay.  As an example of how various signatures may arise, if
the mass ordering is $t>\widetilde{C}_{1}>\tilde{t}>\widetilde{N}_{1}$
and $t>\tilde{t}+\widetilde{N}_{1},$ then a produced $t\bar{t}$ pair
will sometimes (depending on the relative branching ratio, which
depends on the mass values) have one top decay to $W+b $ and the other
to $c+\widetilde {N}_{1},$ giving a $W+2$ jets signature, with the
jets detectable by $b$ or charm tagging, and thus an excess of such events.

$\bullet$ An event was reported by the CDF collaboration from Tevatron
Run 1, $p\bar{p}\rightarrow ee\gamma\gamma$$\noe_T$ , that is
interesting for several reasons, both as a possible signal and to
illustrate some pedagogical issues.  That such an event might be an
early signal of supersymmetry was suggested in 1986.  It can arise
\cite{34,35} if a selectron pair is produced, and if the LSP is
higgsino-like, in which case the decay of the selectron to
$e+\widetilde{N}_{1}$ is suppressed by a factor of $m_{e}.$ Then $\tilde
{e}\rightarrow e+\widetilde{N}_{2}$ dominates, followed by
$\widetilde{N} _{2}\rightarrow\widetilde{N}_{1}+\gamma.$ The only way to
get such an event in the SM is production of $WW\gamma\gamma$ with both
$W\rightarrow e+\nu,$ with an overall probability of order 10$^{-6}$ for
such an event in Run 1.  Other checks on kinematics, cross section for
selectrons, etc., allow a supersymmetry interpretation, and the
resulting values of masses do not imply any that must have been found at
LEP or as other observable channels at the Tevatron, though over some of
the parameter space some associated signal could have been seen.  There
are many consistency conditions that must be checked if such an
interpretation is allowed, and a number of them could have failed but
did not.  Indeed, a related interpretation that had the decay of the
selectron to electron plus very light gravitino is excluded by the
absence of a signal at LEP for events with two photons and large missing
energy.  If this event were a signal additional ones would soon occur in
Run 2.  Because of the needed branching ratios there would be no
trilepton signal at the Tevatron, since $\widetilde{N}_{2}$ decays
mainly into a photon instead of $l^{+}l^{-},$ and the decay of
$\widetilde{N}_{3}$ would be dominated by $\tilde{\nu}\nu.$

Although it might look easy to interpret any non-standard signal or
excess as supersymmetry, in fact a little thought shows it is very
difficult.  As illustrated in the above examples, a given signature
implies an ordering of superpartner masses, which implies a number of
cross section and decay branching ratios.  All must be right.  All
the couplings in the Lagrangian are determined, so there is little
freedom once the masses are fixed by the kinematics of the candidate
events.  To prove a possible signal is indeed consistent with
supersymmetry one has also to check that relations among couplings are
indeed satisfied.  Such checks will be easy at lepton colliders, but
difficult at hadron colliders, so we do not focus on them here.  
There can of course be alternative interpretations of any new physics,
but in all cases it will be possible to show the supersymmetry one is
preferred (if it is indeed correct) --- that is a challenge we would
love to have.

\section{After the first celebration}

Once a signal is found, presumably at the Tevatron, there will of
course be a lot of checking required to confirm it because it will not
be dramatic, as discussed above, but rather excesses in a few channels
that slowly increase with integrated luminosity.  Deducing even the
masses of mass eigenstates may be difficult if more than one channel
contributes significantly to a topological excess.  Nevertheless, it
will be possible to very quickly deduce some general results about
supersymmetry breaking and how it is transmitted.

For example, one of the key questions is the nature of the LSP \cite{36}.
That can immediately exclude some ways to transmit supersymmetry
breaking and favor others, and constrain ideas about how supersymmetry
breaking occurs.  From the discussion above we can make a table whose
columns are various forms the LSP can take and whose rows are
qualitative signatures that do not require complete studies, though
they still require an understanding of the backgrounds:

\begin{center}
\begin{tabular}
[c]{ccccc}
& $\widetilde{B}$ & \~{h} & $\widetilde{G}$ & unstable\\
prompt $\gamma^{\prime}s$ & no & some & yes & no\\
trileptons & yes & no & no & no\\
large $\noe_T$ &  yes & yes & yes & no
\end{tabular}
\end{center}

The table can be extended to other and more detailed LSP descriptions
such as wino LSP, degenerate LSP and NLSP, etc.  It can be extended
to a number of additional signatures and made more quantitative. 
There are some caveats that can be added --- e.g. for the gravitino
case it can happen that long lifetimes for the lightest neutralino
change the signature.  But the basic point that qualitative features
of the excess events will tell us a considerable amount remains.  An
unstable LSP implies that R-parity (or matter parity) is not
conserved, a gravitino LSP implies gauge mediation for the way the
supersymmetry breaking is transmitted, and the bino and higgsino LSP's
suggest gravity mediation, with different consequences for cold dark
matter.

\section{Extensions of the  MSSM}

I want to emphasize that it may be very important to not restrict
analysis of data by over constraining the MSSM with additional
assumptions. Also I have focused on the MSSM for pedagogical simplicity,
but nature could define simplicity differently.  Surely the neutrino
sector must be added, and that affects RGE's for the sectors we have
examined.  There is good motivation for extra U(1) symmetries, which may
lead to extra D-terms and to extra neutralinos that mix to affect the
neutralino mass eigenstates behavior and the CDM physics.  There will be
Planck-scale suppressed operators that may be crucial for flavor physics
and for understanding the fermion masses and for precise calculations of
gauge coupling unification.  There may be extra scalars related to
inflation, and axions, which affect cosmology and CDM physics.  By using
the MSSM without assuming relations among parameters many of these
affects can be allowed for, while if parameters are related by ad hoc
assumptions the extensions could only appear if inconsistencies appeared
in the analysis --- that is hard to see because of the initially large
experimental uncertainties.  For example, extra D-terms shift various
scalar masses and separate $M_{H_{u}}^{2}$ and $M_{H_{d}}^{2},$ so
assuming all the scalars masses are degenerate does not allow the D-term
contributions to appear.

\section{Concluding remarks}

These lectures have emphasized how to construct a supersymmetric
description of nature at the weak scale based on forthcoming data from
colliders, rare decays, static properties, cold dark matter studies,
and more, and how to connect that to a unification scale description,
so that we can eventually learn a complete effective Lagrangian near
the Planck scale.  That is the most that can be achieved by the
traditional approach of science.  If we also understand string theory
(and we do not distinguish here between string theory and M-theory)
well enough, possibly we will be able to bridge the gap to the Planck
scale in 10 dimensions and formulate a fundamental theory.  If so a
number of features of the effective theory will be able to test ideas
about the fundamental theory.  The most important features of the
experimental discovery of supersymmetry will be threefold: we will
understand the natural world better; we will know we are on the right
track to make more progress; and we will be opening a window to see
physics at the Planck scale, which makes immensely more likely that we
will be able to formulate and test a fundamental theory at the Planck
scale.

Sometimes I am asked ``what is left to compute'' by students or
postdocs looking for interesting projects, and interested workers in
related areas.  Much is indeed already known about supersymmetry from
over two decades of work by a number of good people.  But in fact we
have just gotten to the stage where the most important problems can be
addressed$!.$  Little is known about how to relate data to the
parameters of $L_{soft}$ in practice, little is known about the flavor
properties of $L_{soft}$ and how to compute them theoretically or
extract them from data uniquely, and little is known about how to
relate data at the weak scale to an effective Lagrangian at the string
scale.  There is much to understand and to compute.  The third of those
issues will be the main focus of supersymmetry research once
superpartners are being directly studied.

There are several practical features that should be emphasized.
Unless we are missing important basic ideas, a Higgs boson and
superpartners will be produced at the Tevatron collider.
Supersymmetry signals have two escaping LSP's, so they are never
dramatic or obvious or easy to interpret.  They will appear as
excesses in several channels, where channels are labeled by numbers of
leptons and jets, and missing transverse energy.  Once superpartners
are found, the entire challenge to experimenters is to measure the
parameters of $L_{soft},$ which has been the main subject of these
lectures.  The relations of the parameters of $L_{soft}$ to data is
complicated, and it is easy to get the wrong answers if care is not
taken.  Although there seem to be a large number of parameters, any
given measurement depends only on a few, and most parameters enter in
a number of places, so using information from one place in other
analysis will greatly facilitate progress.  Interpreting the data and
learning its implications will be challenging, and it is a challenge
we are eager to have.

\section*{Acknowledgments} 
These lectures have benefited greatly from my collaborations and
discussions with Mike Brhlik, Dan Chung, Lisa Everett, Steve King, and
Lian-Tao Wang.


\begin{thebibliography}{99}
\bibitem{1}A Supersymmetry Primer, S.P.Martin, hep-ph/9709356

\bibitem{2}$\mathit{Perspectives}$ \textit{on Supersymmetry}, G.L.Kane (ed.),
Singapore, World Scientific (1998)

\bibitem{3}I.S.Towner and J.C.Hardy, nucl-th/9809087

\bibitem{4}O.Lebedev and W.Loinaz, hep-ph/0106056; G.Altarelli,
F.Caravaglios, G.F.Giudice, P.Gambino, and G.Ridolfi, hep-ph/0106029;
G.-C.Cho and K.Hagiwara, hep-ph/0105037

\bibitem{5}L.Everett, G.L.Kane, and L.-T. Wang, in preparation

\bibitem{6}S.P.Martin and P.Ramond, Phys.Rev. D51(1995)6515;
hep-ph/9501244; G.Kribs, Nucl. Phys. B535(1998)41; hep-ph/9803259;
K.Dienes, Physics Reports 287 (1997) 447

\bibitem{7}The Supersymmetric Soft-Breaking Lagrangian: Theory and
Applications, D.J.H.Chung, L.Everett, G.L.Kane, S.F.King,
J. Lykken, and L.-T.Wang, in preparation.

\bibitem{8}S.Dimopoulos and H.Georgi, Nucl. Phys. B193(1981)150;
L.Girardello and M.Grisaru, Nucl. Phys. B194(1982)65

\bibitem{9}The literature can be traced from I.Jack and D.R.T.Jones,
Phys. Lett. B457(1999)101; hep-ph/9903365; J.L.Diaz-Cruz, Proceedings
Merida 1999 ``Particles and Fields'', 299.

\bibitem{10}L.Ibanez and G.G.Ross, Nucl. Phys. B368(1992)3; S.P.Martin,
Phys. Rev. D54(1996)2340; hep-ph/9602349

\bibitem{11}L.Ibanez and G.G.Ross, Phys. Lett. B110(1982)215;
L. Alvarez-Gaume, M.Claudsen, and M.Wise, Nucl. Phys. B221(1983)495;
K.Inoue, A.Kakuto, H.Komatsu, and S.Takeshita,
Prog. Theor. Phys. 68(1982)927

\bibitem{12}See H.Haber in \textit{Perspectives on Supersymmetry,
}G.L.Kane (ed.), Singapore, World Scientific (1993)

\bibitem{13}For some review and references see Martin, [1].

\bibitem{14}LEP Electroweak Working Group, LEPEWWG/2001-01

\bibitem{15}G.L.Kane and J.Wells, hep-ph/0003249; see also the
elaboration by M. Peskin and J. Wells, hep-ph/0101342

\bibitem{16}LEP Higgs Working Group for Higgs boson searches
hep-ex/0107029; ALEPH-2001-066; ALEPH Collaboration hep-ph/0201014

\bibitem{17}G.L.Kane, S.F.King, and L.-T. Wang,
Phys. Rev. D64(2001)095013; hep-ph/0010312

\bibitem{18}G.L.Kane and L.-T. Wang, Phys. Lett. B488(2000)383,
hep-ph/0003198; M.Carena, J.Ellis, A.Pilaftsis, and C.Wagner,
Phys. Lett. B495(2000)155, hep-ph/0009212; T.Ibrahim and P.Nath,
Phys. Rev. D63(2001)035009, hep-ph/0008237

\bibitem{19}J.Goldstein, C.Hill, J.Incandela, S.Parke, D.Rainwater, D.Stuart,
Phys. Rev. Lett. 86(2001)1694; L.Reina, S.Dawson, and D.Wackeroth,
hep-ph/0109066; W.Beenakker,S.Dittmaier, M.Kramer, B.Plumper, M.Spira,
P.Zerwas, hep-ph/0107081

\bibitem{20}The literature can be traced from M.Carena, H.Haber,
H.Logan, S.Mrenna, hep-ph/0106116

\bibitem{21}D.Rainwater, D.Zeppenfeld, K.Hagiwara,
Phys. Rev. D59(1999)014037; hep-ph/9808468

\bibitem{22}N.Ghodbane, S.Katsanevas, I.Laktineh, and J.Rosiek, hep-ph/0012031

\bibitem{23}M.Brhlik and G.L.Kane, Phys. Lett. B437(1998)331; hep-ph/9803391

\bibitem{23A} S.Y.Choi, J.Kalinowski, G.Moortgat-Pick, P.M.Zerwas,
hep-ph/0202039 

\bibitem{24}J.Cline, M.Joyce, and K.Kainulainen, hep-ph/0110031

\bibitem{25}M.Carena, J.M.Morena, M.Quiros, M.Seco, and C.Wagner,
Nucl. Phys. B599$\left(  2001\right)  158;$ hep-ph/0011055

\bibitem{26}Lisa L.Everett, G.L.Kane, S.Rigolin, and L.-T. Wang,
Phys. Rev. Lett 86(2001)3484; hep-ph/0102145

\bibitem{27}T.Ibrahim and P.Nath, hep-ph/9908443

\bibitem{28}S.Mrenna, G.L.Kane, and L.-T.Wang,
Phys. Lett. B483(2000)175; hep-ph/9910477

\bibitem{29}M.Brhlik, D.Chung, and G.L.Kane, Int. J. Mod. Phys. D10(2001)367;
hep-ph/0005158

\bibitem{30}M.Brhlik, L.Everett, G.L.Kane, S.F.King, and O.Lebedev,
Phys. Rev. Lett. 84(3041)2000; hep-ph/9909480

\bibitem{31}A.Nelson and L.Randall, Phys. Lett. B316 (1993) 516;
hep-ph/9308277

\bibitem{32}M.Brhlik, L.Everett, G.L.Kane, and J.Lykken,
Phys. Rev. Lett. 83(1999)2124, hep-ph/9905215; Phys. Rev. D62(2000)035005,
hep-ph/9908326

\bibitem{32B}L.Everett, G.L.Kane, S.King, S.Rigolin, and L.-T.Wang,
hep-ph/0202100; See also S.Abel, S.Khalil, and O.Lebedev, hep-ph/0103031
and references therein

\bibitem{33}See chapters by J.-F.Grivaz; M.Carena, R.L.Culbertson,
S.Eno, H.J.Frisch, and S.Mrenna; J.F.Gunion and H.Haber in ref.2.

\bibitem{34}See chapter by G.L.Kane in ref.2.

\bibitem{35}S.Ambrosanio, G.L.Kane, G.Kribs, S.Martin, and S.Mrenna,
Phys. Rev. Lett. 76(1996)3498

\bibitem{36}G.L.Kane, Proceedings of Supersymmetry 1997, ed. M.Cvetic
and P.Langacker.

\bibitem{37}L.Ibanez, C.Munoz, and S.Rigolin, hep-ph/9812397;
Nucl. Phys. B553(1999)43.

\end{thebibliography}
\end{document}